\documentclass[11pt, notitlepage, tightenlines,aps, prd, reprint,superscriptaddress,nofootinbib]{revtex4-1}
\usepackage[utf8]{inputenc}

\usepackage{placeins}
\usepackage{mathtools}
\usepackage{amsfonts}
\usepackage{amssymb,amsmath}
\usepackage{color}
\usepackage[dvipsnames]{xcolor}
\definecolor{lcolor}{rgb}{0.5,0,0}
\definecolor{citcolor}{rgb}{0,0.3,0.0}
\usepackage[breaklinks,colorlinks,urlcolor=blue,citecolor=citcolor,linkcolor=lcolor]{hyperref}
\usepackage{graphicx}
\usepackage{subfigure}
\graphicspath{ {image/} }
\usepackage{multirow}

\newcommand*\diff{\mathop{}\!\mathrm{d}}
\usepackage{braket}
\usepackage{bm}
\usepackage{txfonts}
\usepackage{mathrsfs} 
\usepackage{slashed}
\usepackage{accents}
\usepackage{hepunits}
\usepackage[normalem]{ulem}

\usepackage{soul} 
\usepackage{cancel} 

\usepackage{cleveref}
\crefname{section}{Sec.}{Secs.}
\crefname{figure}{Fig.}{Figs.}
\crefname{appendix}{Appendix}{Appendices}
\crefname{equation}{Eq.}{Eqs.}
\crefname{table}{Table}{Tables}
\Crefname{section}{Section}{Sections}
\Crefname{figure}{Figure}{Figures}
\Crefname{appendix}{Appendix}{Appendices}
\Crefname{equation}{Equation}{Equations}
\Crefname{table}{Table}{Tables}

\usepackage{tikz}
\usetikzlibrary{arrows.meta,calc,decorations.pathmorphing,backgrounds}

\pgfdeclarelayer{background}
\pgfdeclarelayer{foreground}
\pgfsetlayers{background,main,foreground}

\newcommand{\expconfig}[1]{{\langle #1 \rangle_{event}}}
\newcommand{\tr}[1]{\mathrm{tr}\left[ #1 \right]}

\newcommand{\GeV}{{{\,}\textrm{GeV}}}
\newcommand{\fm}{{{\,}\textrm{fm}}}

\begin{document}

\bibliographystyle{apsrev4-1}

\title{Light-front Hamiltonian jet evolution in the Glasma}
\thanks{Authors are listed in alphabetical order}

\author{Dana Avramescu}
\email{dana.d.avramescu@jyu.fi}
\affiliation{Department of Physics, P.O. Box 35, FI-40014 University of Jyv\"{a}skyl\"{a},
Finland}
\affiliation{
Helsinki Institute of Physics, P.O. Box 64, FI-00014 University of Helsinki,
Finland
}

\author{Carlos Lamas}
\email{carloslamas.rodriguez@usc.es}
\affiliation{Instituto Galego de Fisica de Altas Enerxias (IGFAE), Universidade de Santiago de Compostela, E-15782 Galicia, Spain}

\author{Tuomas Lappi}
\email{tuomas.v.v.lappi@jyu.fi}
\affiliation{Department of Physics, P.O. Box 35, FI-40014 University of Jyv\"{a}skyl\"{a},
Finland}
\affiliation{
Helsinki Institute of Physics, P.O. Box 64, FI-00014 University of Helsinki,
Finland
}

\author{Meijian Li}
\email{meijianli@hust.edu.cn}
\affiliation{Instituto Galego de Fisica de Altas Enerxias (IGFAE), Universidade de Santiago de Compostela, E-15782 Galicia, Spain}
\affiliation{School of Physics, Huazhong University of Science and Technology, Wuhan 430074, China}

\author{Carlos A. Salgado}
\email{carlos.salgado@usc.es}
\affiliation{Instituto Galego de Fisica de Altas Enerxias (IGFAE), Universidade de Santiago de Compostela, E-15782 Galicia, Spain}
\affiliation{Axencia Galega de Innovaci\'on (GAIN), Xunta de Galicia, Galicia, Spain}

\begin{abstract}

We develop a light-front Hamiltonian formalism to study the real-time quantum evolution of a high-energy quark propagating through the Glasma phase of a heavy-ion collision. In this work, the quark Fock space is truncated to the $\ket{q}$ sector and the wavefunction is expanded in a discrete basis representation, following the time-dependent Basis Light-Front Quantization (tBLFQ) framework. The classical Glasma background fields enter as a time-dependent external potential, and physical observables are extracted as expectation values of quantum operators over the time-evolved state. We compute the transverse momentum broadening and the jet quenching parameter, finding results consistent with classical estimates, including the expected scaling with respect to the saturation momentum, and use them to perform phenomenological estimations for different collision systems. We also study the color rotation of the quark state induced by the Glasma fields, and examine its dependence on the saturation scale and the gauge choice. This formalism allows systematic improvements to include, in particular, non-eikonal propagation and parton splittings that will be considered in forthcoming publications.

\end{abstract}

\maketitle
\tableofcontents
\section{Introduction}

Jets are collimated sprays of hadrons produced when high-energy quarks and gluons fragment, and they serve as key probes of the quark-gluon plasma (QGP) formed in heavy-ion collisions. As a parton passes through the hot, dense medium, it loses energy and picks up transverse momentum through interactions with the surrounding color fields, an effect known as jet quenching \cite{Gyulassy:1990ye,Casalderrey-Solana:2007knd,Majumder:2010qh,Mehtar-Tani:2013pia,Cao:2020wlm,Apolinario:2022vzg}.
This has been confirmed by a range of measurements at RHIC and the LHC \cite{ATLAS:2010isq,STAR:2016dfv,CMS:2012aa,PHENIX:2001hpc}. While jet quenching in the thermal QGP phase is extensively studied, much less is known about how jets interact with the pre-equilibrium Glasma--an early, short-lived stage dominated by strong color fields \cite{Kovner:1995ja,Lappi:2006fp,Fukushima:2011nq,Gelis:2012ri,Albacete:2014fwa}. Clarifying the role of this phase is important for building a complete picture of jet modification throughout the full evolution of a heavy-ion collision.

The jet quenching parameter $\hat{q}$, which quantifies the transverse momentum squared acquired by a jet per unit path length, is expected to vary with the momentum of the probe and the temperature of the medium, potentially serving as a thermometer for the nuclear matter \cite{Apolinario:2022vzg,JETSCAPE:2021ehl}. Notably, the maximum $\hat{q}$ in the Glasma phase is predicted to be much larger than its value in the QGP \cite{Ipp:2020mjc,Ipp:2021lwz,Carrington:2021dvw,Avramescu:2023qvv}. The large values of $\hat{q}$ reported in the Glasma early stage are compatible with that estimated during the subsequent phase, described using Effective Kinetic Theory (EKT) of QCD \cite{Boguslavski:2023fdm,Boguslavski:2024ezg,Boguslavski:2024jwr,Altenburger:2025iqa,Barata:2025zku}.  The effect of the pre-equilibrium stage on jet quenching and energy loss has been studied within various approaches~\cite{Andres:2019eus,Zigic:2019sth,Adhya:2024nwx,Pablos:2025cli}. These studies have, however, not led to a consensus on the contribution of the pre-equilibrium stage to jet quenching observables. This open puzzle underscores the importance of further investigating jet dynamics in the Glasma.
 
Jet momentum broadening and the quenching parameter in the Glasma fields for eikonal quarks were extracted using a classical  approach in Refs.~\cite{Ipp:2020mjc, Ipp:2020nfu}, using correlators of electric and magnetic fields, where the boost-invariant fields are solved numerically using real-time lattice gauge theory. Comparable values for $\hat{q}$ in the Glasma are obtained using Fokker-Planck transport equations for the quark probe evolving in Glasma fields obtained analytically in the proper time $\tau$ expansion \cite{Carrington:2021dvw,Carrington:2022bnv,Carrington:2026qlg} and in the dilute approximation \cite{Ipp:2020mjc}, or by solving numerically both the Wong transport equations for jets and the classical Yang-Mills equations for the Glasma fields \cite{Avramescu:2023qvv}. Similar calculations were done for heavy quarks \cite{Ruggieri:2018rzi,Sun:2019fud,Carrington:2020sww,Khowal:2021zoo,Ruggieri:2022kxv,Oliva:2024rex,Pooja:2024rnn,Avramescu:2024poa,Avramescu:2024xts,Backfried:2024rub}. These existing approaches are inherently classical in nature, treating the parton as a classical color charge evolving under the Lorentz force of the background Glasma fields, and therefore do not account for quantum effects such as color coherence, spin-field coupling, or the quantum evolution of the parton's internal degrees of freedom. Furthermore, these estimates do not consider radiative processes while the jet propagates in the Glasma. Recently, medium-induced radiation for jets was considered in a model for the Glasma flux tubes \cite{Barata:2024xwy}, where the soft gluon radiation takes place inside correlation domains with fixed length of longitudinal electric fields which are static in time. Notably, the quantum Dirac equation for quark propagating in Glasma background fields has been studied in the context of quark production \cite{Gelis:2004jp,Gelis:2005pb,Gelis:2015eua,Tanji:2017xiw}. However, a quantum treatment of jet propagation in realistic Glasma fields, tracking the full parton quantum state at the amplitude level, remains an open challenge. In this work we will address this situation using the light-front Hamiltonian formalism. 

The light-front Hamiltonian formalism combines light-front dynamics with the Hamiltonian framework (see reviews in \cite{Brodsky:1997de}), providing powerful tools for investigating both the internal structure of bound states (such as Light-Front holographic QCD~\cite{Brodsky:2014yha} and Basis Light-Front Quantization~\cite{Vary:2025yqo} ) and real-time scattering processes~\cite{Zhao:2013cma, Chen:2017uuq, Hu:2019hjx, Lei:2022nsk}. 
The time-dependent Basis Light-Front Quantization (tBLFQ) approach~\cite{Vary:2009gt,Zhao:2013cma} is a non-perturbative computational method built on this formalism. By implementing a basis representation of field quantization, it enables efficient numerical simulations of real-time quantum dynamics. In earlier works, some of us employed tBLFQ to develop a framework for simulating the propagation of quark jets through classical color background fields and carried out systematic studies of jet dynamics in cold nuclear matter and hot dense medium~\cite{Li:2020uhl,Li:2021zaw, Li:2023jeh,Li:2025wzq}. With this framework, one can perform real-time simulations of jets as quantum states at the amplitude level, allowing the investigation of QCD dynamics from first principles and nonperturbative quantum effects that are difficult to access with perturbative approaches.
It is also worth noting recent developments in quantum simulation of jet evolution based on this framework~\cite{ Barata:2022wim,Barata:2023clv,Qian:2024gph,Li:2024ufq}. 

In this work, we apply the light-front Hamiltonian formalism tBLFQ, specifically the jet simulation framework developed in Refs.~\cite{Li:2020uhl,Li:2021zaw, Li:2023jeh,Li:2025wzq}, to study the jet momentum broadening in the initial stages of heavy-ion collisions.
As a first step, we formulate the quark jet in its valence Fock sector and work in the eikonal limit at mid rapidity. We adopt the classical field approach of the Glasma to describe the early-stage medium right after the collision. A key novelty of this work is that the jet is treated as a fully quantum state, whereas previous Glasma studies modeled the jet as a classical probe particle evolving via Wong's equations or Fokker-Planck transport. In our companion work~\cite{Avramescu:2026fgv}, we have elucidated  the distinction between kinetic and canonical momentum and established the connection between the classical and quantum formalism for computing broadening with Glasma fields, which sets the basis of the present study. Here, we derive the corresponding quantum operators for the relevant observables and show that the momentum broadening can be evaluated equivalently via a direct quantum calculation or by integrating the Heisenberg picture quantum operator for the Lorentz force, with both approaches reproducing the classical result within our Fock space truncation. We detail the implementation of the Glasma fields, the coordinate transformation between Milne and jet light-cone variables, and the gauge choices adopted in our formalism, and discuss associated effects such as wave-packet delocalization. We present results for the momentum broadening and jet quenching parameter $\hat{q}$ at mid-rapidity, including its scaling with the saturation scale $Q_s$ and dependence on the infrared regulator, with particular emphasis on the anisotropy between the directions longitudinal and transverse to the collision axis. We also study the color rotation of the quark state induced by the Glasma background fields. 
The quantum results are compared to the results obtained from a classical treatment of the jet. While only the single quark sector is studied here, our work sets up a formalism that can be extended in future works to introduce gluon radiation by including higher Fock states in the theory as well as non-eikonal corrections to the propagation.

This paper is organized as follows. We first introduce the implementation of the Glasma fields and the light-front Hamiltonian formalism for simulating jet evolution in \cref{sec:formalism}. We define the jet observables and derive the relevant quantities in \cref{sec:JetObservables}. We then present and discuss the numerical results in \cref{sec:results}, and conclude in \cref{sec:summary}.
 
\section{Formalism}\label{sec:formalism}

In this section, we present the theoretical framework used to describe the propagation of a quark jet through the Glasma fields. 
We first introduce the physical setup and coordinate systems, followed by the construction of the Glasma background fields and their numerical implementation. We then develop the quantum treatment of the quark jet using the light-front Hamiltonian approach.

\subsection{The setup}
\label{sec:glasma_setup}

\begin{figure} [tb!] 
    \centering
    \includegraphics[width=\columnwidth]{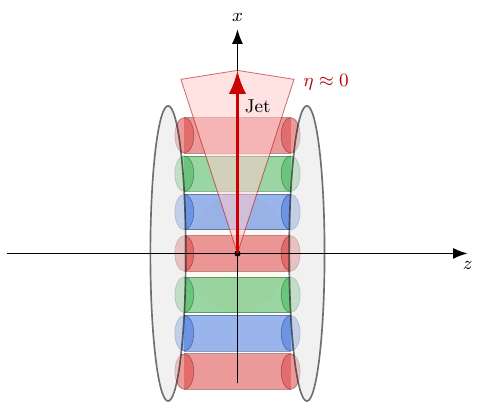}
    \caption{Schematic representation of the setup for our calculation. The colliding high-energy nuclei travel in opposite directions along the $z$-axis, creating correlation domains in the $(x,y)$ plane that are commonly depicted as flux tubes extending in the $z$ direction. For simplicity, these correlation domains are shown at the initial time, each with a length of $1/Q_s$, where $Q_s$ is the saturation scale. The mid-rapidity jet propagates along the $x$-axis, perpendicular to the beam direction, traversing the expanding Glasma fields.}
    \label{fig:Collision}
\end{figure}

In this work, we study the propagation of a mid-rapidity quark jet through the Glasma fields produced in the early phase of a heavy-ion collision. 
The setup of our calculation is illustrated in Fig. \ref{fig:Collision}. The colliding high-energy nuclei propagate in opposite directions along the $z$-axis, while the jet moves along an eikonal trajectory in the $x$-direction, at $\eta\approx 0$. We are interested in how the jet momentum in the $(y,z)$ plane--the transverse plane with respect to the jet propagation direction--is modified by the Glasma fields.

We employ the light-front Hamiltonian approach to evolve the quantum state of the quark in the presence of a classical background external field, in this case the Glasma. A convenient system of coordinates for such calculation is given in terms of light-cone coordinates along the jet direction, denoted as $(x^+, x^-, y, z)$, defined by
\begin{equation}\label{eq:LCcoords}
    x^\pm \equiv t \pm x\,.
\end{equation}
These coordinates differ from the light-cone coordinates of the colliding nuclei with respect to the beam axis, which are denoted as $(x^\bullet,x,y,x^*)$ and defined by
\begin{equation}\label{eq:TLCcoords}
    x^{\bullet} \equiv t + z\, , \qquad x^{*} \equiv t - z\, .
\end{equation}
Here we adopt the notation of Ref.~\cite{Balitsky:1995ub} for the beam axis direction light cone coordinates, to distinguish them from those of the jet direction $x$.

In our setup, the Glasma fields are expressed in Milne coordinates $(\tau, x, y, \eta)$, where $\tau$ denotes the proper time and $\eta$ the space-time rapidity given by
\begin{equation}\label{eq:MilneCoords}
    \tau \equiv \sqrt{t^2 - z^2},\quad \eta \equiv \tanh^{-1}\left(\frac{z}{t}\right)\,.
\end{equation}
This coordinate system is particularly suitable to describe boost-invariant fields, which do not depend on the rapidity $\eta$. In practice, the jet evolution is computed in light-cone coordinates, while the Glasma fields are constructed in Milne coordinates and subsequently mapped onto the light-cone components needed for the Hamiltonian evolution.
Further details on the coordinate systems and conventions used in this work are provided in \cref{app:Coordinate_Conventions}.

\subsection{The Glasma fields}

\subsubsection{The boost-invariant Glasma} \label{sec:GlasmaFields}

The Glasma fields, which serve as the external background field in our calculation, are obtained as follows; further details can be found in \cite{Avramescu:2026fgv}. The Glasma consists of classical gluon fields obtained from the collision of high-energy nuclei within the Color Glass Condensate framework \cite{Iancu:2003xm,Gelis:2010nm,Gelis:2012ri}. At such energies, the nucleus is dominated by large numbers of soft gluons, which carry a small fraction of the total momentum. The hard partons inside the nucleus, treated as a classical color current $\mathcal{J}^\mu$, radiate the soft partons, described by a classical gauge field $\mathcal{A}^\mu$. The evolution of the latter is governed by the classical Yang-Mills equation 
\begin{equation} \label{eq:CYM}
    [\mathcal{D}_\mu,\mathcal{F}^{\mu\nu}]=\mathcal{J}^\nu,\,
\end{equation}
where $[\mathcal{D}_\mu,\chi]=\partial_\mu\chi+ig[\mathcal{A}_\mu,\chi]$ is the covariant derivative and $\mathcal{F}^{\mu\nu}=\partial^\mu\mathcal{A}^\nu-\partial^\nu\mathcal{A}^\mu+ig[\mathcal{A}^\mu,\mathcal{A}^\nu]$ the field strength tensor.  

Before the collision, the two nuclei, denoted as $1$ and $2$, propagate along the light-cone directions $x^\bullet$ and $x^*$, respectively, with the nuclei light-cone coordinates defined in \cref{eq:TLCcoords}. The associated color current is aligned with the corresponding light-cone direction $x^{\bullet, *}$ and encodes the transverse color charge density: $\mathcal{J}^\mu_{(1,2)}\propto\delta^{\mu,(\bullet, *)}\delta(x^{*, \bullet})\rho_{(1,2)}(\vec{x}_\perp)$ where $\vec{x}_\perp\equiv(x,y)$. By the uncertainty principle, the highly boosted sources are effectively static in the conjugate light-cone coordinate, so that 
$\rho_{(1,2)}(x^{*, \bullet},\vec{x}_\perp)\approx \delta(x^{*, \bullet})\rho_{(1,2)}(\vec{x}_\perp)$. For large nuclei at high-energies, the color charges are sampled from a Gaussian probability functional using the McLerran-Venugopalan (MV) model \cite{McLerran:1993ni, McLerran:1993ka, McLerran:1994vd}. This enforces overall color neutrality via $\expconfig{ \rho^a(\vec{x}_\perp)}=0$ and yields the $2$-point charge correlation
\begin{equation} \label{eq:MVmodel}
   \expconfig{\rho^a(\vec{x}_\perp)\,\rho^b(\vec{y}_\perp)}=g^2\mu^2 \,\delta^{ab}\delta^{(2)}(\vec{x}_\perp-\vec{y}_\perp),\,
\end{equation}
where $\expconfig{\dots}$ denotes the average over multiple event configurations and $g^2\mu$ is the color charge density of the individual nuclei $1$ and $2$. The color charge parameter $g^2\mu$ is directly related to the saturation momentum $Q_s$, the scale at which the gluon distribution saturates \cite{Albacete:2014fwa,Lappi:2007ku}. In practice, to preserve randomness in the infinitesimal direction of propagation $x^\bullet(x^*)$ \cite{Fukushima:2007ki}, one retains the dependence $\rho_{(1,2)}(x^{*, \bullet},\vec{x}_\perp)$ and the MV model correlator from \cref{eq:MVmodel} is regularized as
\begin{equation}\label{eq:MVModelSheets}
    \expconfig{\rho^a_m(\vec{x}_\perp)\,\rho^b_n(\vec{y}_\perp)}=\frac{\delta_{m,n}}{N_\eta}\,g^2\mu^2 \,\delta^{ab}\delta^{(2)}(\vec{x}_\perp-\vec{y}_\perp),\,
\end{equation}
in which $N_\eta$ denotes the total number of layers of color sources along $x^\bullet(x^*)$, and the layer indices $m,n$ run over $1,2,\cdots, N_\eta$.
We start by obtaining the single-nucleus gauge fields, needed to construct the Glasma initial condition. In the covariant gauge $\partial_\mu\mathcal{A}^\mu_\mathrm{cov}=0$, the classical Yang-Mills equation \eqref{eq:CYM} reduces to a two-dimensional Poisson equation 
\begin{equation} \label{eq:Poisson}
    (\Delta_\perp-m_g^2)\,\mathcal{A}^{\bullet, *}_\mathrm{cov}(x^{*, \bullet},\vec{x}_\perp)=-\rho_{(1,2)}(x^{*, \bullet},\vec{x}_\perp)\,,
\end{equation}
where $\mathcal{A}^\bullet_\mathrm{cov}$ and $\mathcal{A}^*_\mathrm{cov}$ are the gauge fields sourced by nucleus 1 (traveling along $x^\bullet$) and nucleus 2 (traveling along $x^*$). Here, $\Delta_\perp$ is the transverse Laplace operator and $m_g$ an infrared regulator. In practice, \cref{eq:Poisson} is solved in momentum space via Fourier transformation. 
One then performs a gauge transformation to the nuclei light-cone gauge $\mathcal{A}_\mathrm{LC}^{\bullet, *}=0$, given by the color rotation
\begin{equation}
    \mathcal{V}_{(1,2)}(\vec{x}_\perp)=\mathcal{P}\exp\left[\int \diff x^{*, \bullet} \mathcal{A}^{\bullet, *}_\mathrm{cov}(x^{*,\bullet},\vec{x}_\perp)\right]\;,
\end{equation}
yielding pure gauge fields for each nucleus: 
\begin{equation}\label{eq:PureGauge}
    \alpha_{(1,2)}^{\underline{i}}(\vec{x}_\perp)=\frac{i}{g}\mathcal{V}_{(1,2)}(\vec{x}_\perp)\partial^{\underline{i}}\mathcal{V}_{(1,2)}^\dagger(\vec{x}_\perp)\,.
\end{equation}
Here, $\alpha_{(1,2)}^{\underline{i}}\equiv \mathcal{A}^{\underline{i},\mathrm{LC}}_{(1,2)}$ denotes the transverse components with $\underline{i}\in \{ x,y \}$. 

These pure gauge fields serve as the building blocks for the Glasma initial condition at $\tau=0$. The standard Glasma initial condition is formulated in Milne coordinates, with proper time $\tau$ and space-time rapidity $\eta$ defined in \cref{eq:MilneCoords}, in the temporal gauge $\mathcal{A}_\tau=0$ and under the assumption of boost invariance $\partial_\eta\mathcal{A}_\mu=0$. At $\tau=0$, the Glasma fields are expressed in terms of the pure gauge fields 
\cite{Kovner:1995ts,Krasnitz:1998ns},
\begin{subequations} \label{eq:GlasmaInitialCondition}
    \begin{align} 
        &\mathcal{A}^{\underline{i}} (\tau = 0) = \alpha_{(1)}^{\underline{i}} + \alpha_{(2)}^{\underline{i}}\, , \\
        &\mathcal{A}^\eta (\tau = 0) = \frac{ig}{2}[\alpha_{(1)}^{\underline{i}}, \alpha_{(2)}^{\underline{i}}]\, ,
    \end{align}
\end{subequations}
and the temporal derivatives are imposed as $\partial_\tau \mathcal{A}^i = \partial_\tau \mathcal{A}^\eta = 0$. The subsequent time evolution of the Glasma fields is governed by the sourceless classical Yang-Mills equation $[\mathcal{D}_\mu,\mathcal{F}^{\mu\nu}]=0$, which for the boost-invariant Glasma in the temporal gauge reduces to
\begin{subequations} \label{eq:CYMMomenta}
    \begin{align}
        &\partial_\tau \,(\tau\,\partial_\tau\mathcal{A}_{\underline{i}}) = \tau\, \mathcal{D}_{\underline{j}} \mathcal{F}_{\underline{j}\underline{i}}-\frac{ig}{\tau}\,[\mathcal{A}_\eta,\mathcal{D}_{\underline{i}}\mathcal{A}_\eta]\,,\\
        &\partial_\tau \,\left(\frac{1}{\tau}\,\partial_\tau\mathcal{A}_\eta\right) = \frac{1}{\tau}\,\mathcal{D}_{\underline{i}}(\mathcal{D}_{\underline{i}}\mathcal{A}_\eta)\;.
    \end{align}
\end{subequations}

The field equations \eqref{eq:CYMMomenta} are solved numerically using real-time lattice gauge theory. The transverse plane $(x,y)$ of side length $L_\perp$ is discretized into $N_\perp$ lattice points along each direction, giving a lattice spacing $a_\perp=L_\perp/N_\perp$. The transverse gauge field $\mathcal{A}_{\underline{i}}$ is encoded in gauge link variables
\begin{equation}\label{eq:GaugeLink}
    U_{\underline{i}}\,(\tau,\vec{x}_\perp)= \exp\left[ig a_\perp \mathcal{A}_{\underline{i}}\left(\tau,\vec{x}_\perp+\frac{a_\perp}{2}\vec{e}_{\underline{i}}\right)\right]\,,
\end{equation}
where $\vec{e}_{\underline{i}}$ is the unit vector along direction $\underline{i}$. These gauge links represent the smallest Wilson lines connecting neighboring lattice sites and are used to construct plaquette variables, corresponding to the shortest Wilson loops on the transverse lattice. Due to boost invariance, the field component $\mathcal{A}_\eta$ is not discretized using such variables. The field equations
\cref{eq:CYMMomenta} are then expressed in terms of $U_{\underline{i}}$, $\mathcal{A}_\eta$, and their time derivatives, and solved numerically. 
Further details on the numerical implementation for SU($3$) Glasma fields are provided in \cite{Muller:2019bwd,Ipp:2020mjc}. 
 
The Glasma electric and magnetic fields, extracted from the gauge potentials as 
\begin{align}\label{Eq:EBFieldsGlasmaClass}
    \begin{split}
           &  \mathcal{E}_{\underline{i}} =\partial_\tau\mathcal{A}_{\underline{i}}\,,
            \qquad
         \mathcal{E}_z  =  \frac{1}{\tau\,}\partial_\tau \mathcal{A}_\eta\,, \\
           & \mathcal{B}_{\underline{i}}  = \frac{1}{\tau} \varepsilon_{\underline{i}\underline{j}} \,\mathcal{D}_{\underline{j}}\mathcal{A}_\eta,
             \qquad
             \mathcal{B}_z  = -\mathcal{F}_{xy}\, ,
    \end{split}
\end{align}
exhibit characteristic correlation patterns. At the initial time $\tau=0$, the longitudinal electric fields $\mathcal{E}_z$ are correlated on a typical scale $1/Q_s$, corresponding to the correlation domain depicted in \cref{fig:Collision}, while the longitudinal magnetic fields $\mathcal{B}_z$ also exhibit an anti-correlation region \cite{Dumitru:2013wca,Dumitru:2014nka,Ipp:2020mjc,Ruggieri:2017ioa}. 

To reduce numerical errors in the computation of physical observables, it is advantageous to choose a gauge in which the magnitude of the gauge potential is minimized \cite{Avramescu:2026fgv}. The residual gauge freedom of the Glasma temporal gauge $\mathcal{A}_\tau=0$ can be used to additionally impose the transverse Coulomb gauge condition
\begin{equation}
    \sum_{\underline{i}}\partial_{\underline{i}}\mathcal{A}_{\underline{i}}(\tau,\vec{x}_\perp)=0\,,
\end{equation}
at the initial time $\tau=0$, as done in \cite{Lappi:2003bi,Blaizot:2010kh,Boguslavski:2018beu}. Such a gauge fixing minimizes the square of the transverse gauge potential $\vec{\mathcal{A}}_\perp\equiv (\mathcal{A}_x, \mathcal{A}_y)$, namely $\int \diff \vec{x}_\perp \vec{\mathcal{A}}_\perp^2(\vec{x}_\perp)$ at a given $\tau$ \cite{Gubarev:2000eu,Gubarev:2000nz}. The Coulomb gauge is imposed numerically using the Fourier acceleration technique \cite{Davies:1987vs,Cucchieri:1995pn}, as employed in \cite{Avramescu:2026fgv}. The resulting gauge transformation $\mathcal{C}(\tau=0,\vec{x}_\perp)\equiv \mathcal{C}(\vec{x}_\perp)$ is then applied to the Glasma transverse gauge links $U_{\underline{i}}(\vec{x}_\perp)\rightarrow\mathcal{C}(\vec{x}_\perp)\,U_{\underline{i}}(\vec{x}_\perp)\,\mathcal{C}^\dagger(\vec{x}_\perp+a_\perp\vec{e}_{\underline{i}})$ and fields $\mathcal{A}_\eta(\vec{x}_\perp)\rightarrow\mathcal{C}(\vec{x}_\perp)\,\mathcal{A}_\eta(\vec{x}_\perp)\,\mathcal{C}^\dagger(\vec{x}_\perp)$. As shown in \cite{Avramescu:2026fgv}, working in the Coulomb gauge significantly reduces numerical errors arising from the discretization.

In practice, any physical quantity $\mathcal{O}[\rho]$ extracted from the numerical simulations is computed event-by-event, where each event is characterized by a color charge configuration $\rho$ sampled according to the MV model \eqref{eq:MVmodel} with a fixed value of $g^2\mu$. The configuration average over $N_\text{events}$ events is defined as 
\begin{equation} \label{eq:ConfigAvg}
    \expconfig{\mathcal{O}[\rho]}=\frac{1}{N_{events}}\sum_{1}^{N_{events}} \mathcal{O}[\rho]\,.
\end{equation}
The proportionality factor between $g^2\mu$ and $Q_s$ is determined from the dipole correlator of Wilson lines before the collision, and depends on the numerical parameters entering the Glasma fields, notably on the infrared regulator $m_g$ and the lattice spacing $a_\perp$ \cite{Lappi:2007ku}. 

\subsubsection{Classical jets in Glasma} \label{sec:ClassicPropagation}

Before developing the quantum treatment in \cref{sec:tBLFQ}, we briefly review the classical formulation of jet propagation in a non-Abelian background field. This establishes key physical quantities---in particular the Lorentz force---and the notation used throughout the paper, and allows for direct comparison with existing results in the literature \cite{Ipp:2020mjc,Ipp:2020nfu,Avramescu:2023qvv,Avramescu:2026fgv}.

The classical evolution is performed by using Wong's equations \cite{Wong:1970fu}, which represent the equations of motion for classical point-like particles in the presence of a non-Abelian background field. The classical transport equation for the kinetic momentum $p_{kin}^\mu$ is
\begin{equation}
    \frac{\diff p^\mu_{kin}}{\diff \lambda}=\frac{g}{T_R} \tr{Q(\lambda) \,f^\mu(\lambda)}\,,
\end{equation}
where $\lambda$ is the affine parameter which parametrizes the particle worldline, $f^\mu$ the Lorentz force along the particle trajectory and $Q=Q^a t^a$ the classical color charge. Here $t^a$ are the generators of the gauge group SU($3$) and $\mathrm{tr}\left[t^a t^b\right]=T_R\delta^{ab}$ with $T_F=1/2$ for $R=F$ quarks in the fundamental representation. The classical Lorentz force is defined as
\begin{equation}\label{eq:f_classical}
    f_\mu\equiv\frac{\diff x^\nu}{\diff \lambda}\mathcal{F}_{\mu\nu}\,,
\end{equation}
where $x^\mu$ is the particle coordinate and $\mathcal{F}^{\mu\nu}=\mathcal{F}^{\mu\nu,a}t^a$ contains the color electric and magnetic fields experienced by the particle. This is accompanied by an equation of motion for the classical color charge, which is expressed as a color rotation
\begin{equation}
    Q(\lambda)=\mathcal{U}(\lambda;0)\,Q(0)\,\mathcal{U}^\dagger(\lambda;0)\,,
\end{equation}
with a Wilson line constructed along the particle trajectory which accumulates the background gauge field as
\begin{equation}\label{eq:PartWilsonLine}
    \mathcal{U}(\lambda;0)=\mathcal{P}\exp\left[-ig\int_0^\lambda \diff\lambda^\prime \frac{\diff x^\mu}{\diff\lambda^\prime} \mathcal{A}_\mu(\lambda^\prime)\right].
\end{equation}

We study classical momentum broadening, which is defined as the squared difference in the kinetic momentum along directions $i\in(x,y,z)$ as
\begin{equation}
    \delta p_i^{kin}(\lambda)\equiv p_i^{kin}(\lambda)-p_i^{kin}(0)\,.
\end{equation}
Further, we focus on the propagation of an eikonal quark jet in the Glasma background fields that were described above in \cref{sec:GlasmaFields}. We consider the quark to move with a large momentum $p^x\rightarrow\infty$ on a straight-line trajectory along the $x$-axis, that is $x=t$ along with $\lambda=\tau$ at mid-rapidity $\eta=0$. In this limiting case, one may integrate over the color charge degrees of freedom using the classical color averages \cite{Kelly:1994dh,Litim:1999id,Litim:2001db,Carrington:2016mhd,Ipp:2020mjc,Avramescu:2023qvv}
\begin{equation}\label{eq:intqaqb}
    \int\diff Q\,Q^aQ^b=T_R\delta^{ab}\,.
\end{equation}
In general, classical color charge averages are done as
\begin{equation}\label{eq:defintq}
    \langle X\rangle_Q\equiv\int \diff Q\,X\,,
\end{equation}
with $X$ any $n$-point function of $Q^a$. The color charged averaged kinetic momentum broadening experienced by the quark jet is given by 
\begin{equation}\label{eq:KinMomBroadJet}
    \langle (\delta p_i^{kin}(\tau))^2\rangle_Q=g^2\int_0^\tau \diff \tau^\prime \int_0^\tau \diff\tau^{\prime \prime}\,\tr{ \tilde{f}_i(\tau^\prime)\tilde{f}_i(\tau^{\prime \prime})}\,,
\end{equation}
where the Lorentz force is color rotated with the particle Wilson line
\begin{equation} \label{eq:ClassColorRotLorentz}
    \tilde{f}_i(\tau)=\mathcal{U}(\tau;0)\,f_i(\tau)\,\mathcal{U}^\dagger(\tau;0)\,.
\end{equation}
For eikonal jets in temporal gauge, the Wilson line from \cref{eq:PartWilsonLine} simplifies to
\begin{equation}
    \mathcal{U}(\tau;0)=\mathcal{P}\exp\left[-ig\int_0^\tau \diff \tau^\prime \,\mathcal{A}_x(\tau^\prime)\right]\,.
\end{equation}
The Lorentz force exerted on the quark jet along the $i\in(y,z)$ directions contains the following combination of Glasma electric and magnetic fields
\begin{subequations}\label{eq:JetLorentzForce}
    \begin{align} 
        &f^y=\mathcal{E}_y-\mathcal{B}_z\,,\\ &f^z=\mathcal{E}_z+\mathcal{B}_y\,.
    \end{align}
\end{subequations}
The resulting kinetic momentum broadening, as expressed in \cref{eq:KinMomBroadJet}, is thus evaluated as the correlator of Glasma electric and magnetic fields along the eikonal jet trajectory \cite{Ipp:2020mjc,Ipp:2020nfu,Avramescu:2023qvv,Avramescu:2026fgv}. We emphasize that  there is a difference between the kinetic and canonical momentum broadening, with the canonical broadening being gauge dependent, as discussed in more detail in \cite{Avramescu:2026fgv}.

\subsubsection{Glasma fields on the jet light-cone}\label{sec:Glasma_jetLC}

To describe the evolution of the quantum quark state we will use the light-front Hamiltonian formalism, which is naturally expressed in the light cone coordinates of the jet, introduced in \cref{eq:LCcoords}. The Glasma fields, which are derived and evolved in Milne coordinates, must then be re-written into this coordinate system. 

We focus on high-energy quark jets propagating along eikonal trajectories with $x =t $. Along such a trajectory, one has $x(x^+) = x^+/2$, and the corresponding proper time is given by $\tau(x^+,z) = \sqrt{(x^+/2)^2-z^2}$. It follows that the relevant components of the Glasma fields experienced by the quark jet are
\begin{equation}\label{eq:ExactJetTransformation}
    \begin{split}
        \mathcal{A}_+ (x^+, y, z) &= \frac{1}{2} \left[\mathcal{A}_x \left(\tau(x^+, z), x(x^+), y \right) \right. \\
        & \hspace {40 pt} \left. -\frac{z}{\tau^2}\mathcal{A}_\eta \left(\tau(x^+, z), x(x^+), y \right) \right]\,, \\
        \mathcal{A}_y (x^+, y, z) &= \mathcal{A}_y\left(\tau(x^+, z), x(x^+), y \right)\,, \\
        \mathcal{A}_z (x^+, y, z) &= \frac{x^+}{2\tau^2}
        \mathcal{A}_\eta \left(\tau(x^+, z), x(x^+), y \right)\;.
    \end{split}
\end{equation}
On the lattice, the transverse gauge potentials $\mathcal{A}_{\underline{i}}$ with $\underline{i}\in \{x,y\}$ are extracted as the logarithm of the corresponding gauge links, that is $iga_\perp\mathcal{A}_{\underline{i}}=\ln U_{\underline{i}}$ using \cref{eq:GaugeLink}. 

The numerical implementation of the exact coordinate transformation between Milne and light-cone variables presents several challenges. For non-vanishing rapidity, one has $t \neq \tau$ and the Glasma fields at a given $x^+$ value may contain contributions from all previous proper time values $\tau < x^+/2$. Consequently, one would need to retain the full history in Milne proper time $\tau$ to reconstruct the fields at a fixed light-cone time $x^+$, which is computationally costly. Note that the Glasma field equations from \cref{eq:CYMMomenta} contain only first order proper time derivatives and the numerical implementation of these equations does not require fully storing the fields at all previous $\tau$ values. In addition at a fixed $t$, the classical background fields occupy a finite extent in the $z$ direction, $\Delta z = t$, whereas the quantum jet is described by a spatially extended wave packet rather than a pointlike particle. Consequently, parts of the quark wave function may extend beyond the Glasma light-cone and probe regions where the classical fields are not physically defined. This reflects an intrinsic limitation of describing the state of a quantum particle, with $t$ or $x^+$ as the time coordinate, at an early time after the collision. At early times in particular, the extent of the collision region in $z$ is very small, and it is not straightforward to consistently account for the finite spatial extent of the wavefunction of the jet.

These complications associated with the coordinate transformation are mitigated by exploiting the fact that the jet is well localized around mid-rapidity. In the classical calculation ~\cite{Ipp:2020mjc}, only the first-order derivatives in $z$ contribute to the momentum broadening from \cref{eq:KinMomBroadJet}, as may be noticed from the expressions of the relevant electric and magnetic field components, see \cref{Eq:EBFieldsGlasmaClass}. 
This suggests that a first-order expansion in the $z$ coordinate, $\mathcal{A}_\mu (x^+, y, z) \simeq \mathcal{A}_\mu (x^+, y, 0) + $ $(z\, \partial \mathcal{A}_\mu/\partial z)(x^+, y, 0)+\mathcal{O}(z^2)$, should already capture the relevant physics for jets propagating near $\eta \simeq 0$. The only component that has a non-trivial contribution from the linear term is $\mathcal{A}_+$, while for $\mathcal{A}_y$ and $\mathcal{A}_z$ the linear term vanishes and only the zero order term contributes. The expanded fields then read
\begin{align} \label{eq:ApproxJetTransformation}
    \begin{split}
        \mathcal{A}_+ (x^+, y, z) & \simeq \frac{1}{2} \left[\mathcal{A}_x \left(\tau(x^+), x(x^+), y \right) \right. \\
        & \hspace{40 pt} \left. -\frac{4 z}{(x^+)^2}\mathcal{A}_\eta \left(\tau(x^+), x(x^+), y \right) \right]\,, \\
        \mathcal{A}_y (x^+, y, z) & \simeq \mathcal{A}_y\left(\tau(x^+), x(x^+), y \right)\,, \\
        \mathcal{A}_z (x^+, y, z) & \simeq \frac{2}{x^+}
        \mathcal{A}_\eta \left(\tau(x^+), x(x^+), y \right)\,,
    \end{split}
\end{align}
where $\tau(x^+) = x(x^+) = x^+/2$. Using this expansion we effectively extend the fields outside the future light cone in a smooth way, preserving the dynamics in the region around $\eta=0$ that is our main interest here. Now the only dependence on $z$ is given by the $\mathcal{A}_\eta$ term of $\mathcal{A}_+$. However, this approximation only works for jet wavefunctions which are well localized around mid-rapidity and breaks if the effective transverse width of the jet becomes too large. From now on we simply denote
\begin{equation}\label{eq:notamuxplus}
    \mathcal{A}_\mu \left(\tau(x^+), x(x^+), y \right)\equiv \mathcal{A}_\mu \left(x^+\right)\,,
\end{equation} 
since the $x$ coordinate coincides with the temporal one and the $y$ coordinate is not modified by the coordinate transformation.

\subsection{Jet evolution in the light-front Hamiltonian approach}\label{sec:tBLFQ}

Let us now summarize the formalism used for the quantum evolution of a quark state in the presence of a classical non-Abelian background field. We start by deriving the Hamiltonian using the framework of light-front quantization. We then present the method employed to construct the quark state and explain how the state is numerically evolved. Note that here, both the fields and the momenta must be understood as quantum operators acting on the quark state. 

\subsubsection{The quark propagation Hamiltonian}

The Lagrangian that describes the evolution of the quark fermion field ($\Psi$) in the presence of a classical external gluon field ($\mathcal{A}_\mu$) can be derived from the Dirac Lagrangian using minimal coupling, where the covariant derivative is modified as $\partial_\mu \to \mathcal{D}_\mu = \partial_\mu +ig \mathcal{A}_\mu$ with $\mathcal{A}_\mu = \mathcal{A}_\mu^a t^a$. As an initial investigation, we truncate the Fock space of the quark jet $\ket{q}+\ket{qg}+\ket{qgg}+\cdots$ to the single-quark sector $\ket{q}$, thereby neglecting contributions from gluon emissions and virtual corrections associated with the quantum gluon field. The Lagrangian is then given by
\begin{align}\label{eq:Lagrangian}
    \mathcal{L} = \overline{\Psi} (i\gamma^\mu \mathcal D_\mu - m) \Psi\;.
\end{align}

The corresponding Hamiltonian is obtained from the Lagrangian in \cref{eq:Lagrangian} by performing a Legendre transform. We work in the light-front formalism, where $x^+$ defined in \cref{eq:LCcoords} serves as the temporal variable. The resulting Hamiltonian is referred to as the light-front (LF) Hamiltonian \cite{Brodsky:1997de}. It takes the form 
\begin{align}\label{eq:LFHamiltonian}
    &P^-(x^+) = \int \diff x^- \diff^2\vec{x}_\perp \Bigg[
    \frac{1}{2} \bar{\Psi}\gamma^+ 
    \frac{m^2-\nabla_\perp^2}{i\partial^+-g \mathcal A^+} \Psi \\
    &+ g \bar{\Psi} (\gamma^+ \mathcal{A}_+ + \gamma^i \mathcal{A}_i) \Psi + \frac{g^2}{2} \bar{\Psi}\gamma^i\mathcal{A}_i 
    \frac{\gamma^+}{i\partial^+-g \mathcal A^+} 
    \gamma^j \mathcal{A}_j \Psi
    \Bigg]\notag\,.
\end{align}
We use lower indices for the field $\mathcal{A}_\mu$, with the exception of the term $i\partial^+-g \mathcal A^+$, where we retain the upper index to make transparent the physical interpretation of $p^+-g \mathcal{A}^+$ as the kinetic longitudinal momentum. 

Light-front quantization of the gluon field is formulated in the light-cone gauge $A^+=0$. However, since the quantized gluon field is absent due to Fock space truncation, only the classical external field $\mathcal{A}_\mu$ appears in the resulting Hamiltonian. We emphasize that the external classical field $ \mathcal{A}_\mu$ can be in a different gauge, and thus the component $\mathcal A^+$ still appears. In practice, as discussed in \cref{sec:GlasmaFields} and done in \cite{Avramescu:2026fgv}, the Glasma fields are obtained in the Fock-Schwinger gauge $\mathcal{A}_\tau = 0$, and we additionally impose the transverse Coulomb gauge $\partial_{\underline{i}}\mathcal{A}_{\underline{i}}=0$ at the initial proper time $\tau=0$.

Formally, the derivative in the denominators of \cref{eq:LFHamiltonian} may be understood as $i\partial^+ \to p^+$, which is valid when acting on momentum eigenstates where $p^+$ is treated as a c-number. In this work, we only consider eikonal jets whose longitudinal momentum is very large $p^+\rightarrow\infty$. We then expand \cref{eq:LFHamiltonian} assuming $p^+\gg \mathcal{A}^+$, keeping terms of order $1/p^+$ but neglecting ones of order $\mathcal{A}^+/(p^+)^2$, to get
\begin{multline}\label{eq:LFHamiltonianSubEik}
    P^-(x^+) = \int \diff x^- \diff^2\vec{x}_\perp \Bigg[
    \frac{1}{2} \bar{\Psi}\gamma^+ 
    \frac{m^2-\nabla_\perp^2}{p^+} \Psi \\
    + g \bar{\Psi} (\gamma^+ \mathcal{A}_+ + \gamma^i \mathcal{A}_i) \Psi + \frac{g^2}{2} \bar{\Psi}\gamma^i\mathcal{A}_i 
    \frac{\gamma^+}{p^+} 
    \gamma^j \mathcal{A}_j \Psi
    \Bigg]\,.
\end{multline}
Therefore, for an arbitrary gauge with finite $\mathcal{A}^+$, the longitudinal $\mathcal{A}^+$ component of the background field does not enter the Hamiltonian at $\mathcal{O}(1/p^+)$ in the eikonal expansion. We will, however, keep the terms suppressed as $1/p^+$ since they will be relevant in the derivation of quantum operators.

We quantize the fermion field by expanding the field operator in terms of creation and annihilation operators,
\begin{align}\label{eq:free_mode_expansion}
    \begin{split}
        \Psi_c(x)=&\sum_{\lambda } \int\frac{\diff p^+\diff^2 \vec{p}_\perp }{\sqrt{2 p^+ (2\pi)^3}}   
        \\ &\times[ u_\lambda(p) e^{-ip\cdot x}b_{\lambda,c}(p) +  v_\lambda(p) e^{ip\cdot x}d^\dagger_{\lambda,c}(p)]\, ,
    \end{split}
\end{align}
where $p\cdot x=1/2p^+ x^- -\vec p_\perp\cdot \vec x_\perp$ denotes the light-front spatial three-product, $\lambda$ the light-front helicity, and $c$ the color index and $u_\lambda$ and $v_\lambda$ the Dirac spinors. The phase factor originates from the four-product
$p\cdot x=\frac{1}{2}(p^+x^-+p^-x^+)-\vec p_\perp\cdot\vec x_\perp$, which reduces to the above form upon quantization on the $x^+=0$ surface. This reduction is a feature of canonical quantization in the Schr\"{o}dinger picture, as implemented here in tBLFQ. The evolution in $x^+$ is then generated by the Hamiltonian acting on the state, while equal-$x^+$ anti-commutation relations 
\begin{align}
    \begin{split}
        \{b_{\lambda,c}(p),b^{\dagger}_{\lambda',c'}(p')\}
        &=\{d_{\lambda,c}(p),d^{\dagger}_{\lambda',c'}(p')\}\\
        &=\delta^{(3)}(p-p')\,\delta_{\lambda,\lambda'}\,\delta_{c,c'}\,.
    \end{split}
\end{align}
remain independent of the chosen picture. Note that within our leading Fock space truncation only the term containing the quark spinor contributes. Applying this free-field expansion to the Hamiltonian in \cref{eq:LFHamiltonianSubEik} and presenting the result in the quantum many-body representation (see \cref{app:HamiltonianSimplification} for detailed derivations) leads to
\begin{align} \label{eq:1stQuantHamiltonian}
   P^- = \frac{m^2 + (\vec{p}_\perp -g\vec{\mathcal{A}}_\perp)^2}{p^+} + 2g\mathcal{A}_+ - 2g\frac{\mathcal{B}_x S^x}{p^+}  \, .
\end{align}
Our derivation closely follows the framework introduced in Ref.~\cite{Wen:2025dwy} for the light-front Hamiltonian of quarkonium in an external magnetic field.
This many-body representation provides an intuitive understanding and facilitates the derivation of various operator expressions.

The first term in \cref{eq:1stQuantHamiltonian} is the kinetic energy of the quark propagating in the presence of the background field, from which we identify the kinetic momentum 
\begin{equation} \label{eq:KineticMomentum}
    \vec p^{\, kin}_\perp= \vec{p}_\perp -g\vec{\mathcal{A}}_\perp\,,
\end{equation}
where $\vec{p}_\perp$ is the canonical momentum conjugated to the transverse position operator. A more detailed discussion of the distinction between the kinetic and canonical momentum in the Glasma is provided in our companion paper~\cite{Avramescu:2026fgv}. Note that in light-front dynamics, $p^+$ is the central charge of the transverse Galilei subgroup and plays the role of an effective mass in the transverse motion. It is therefore natural that the resulting Hamiltonian resembles the quadratic form of the color Dirac Hamiltonian, e.g., as given in Ref.~\cite{Heinz:1984yq}.

The second term in \cref{eq:1stQuantHamiltonian} describes the interaction of the quark with the external potential. The last term accounts for the coupling of the quark's spin projection to the chromo-magnetic field $ \vec{\mathcal{B}}=\nabla \times \mathcal{\vec A}$, analogous to the Zeeman effect in the non-relativistic (NR) quantum mechanics. Recall that in the quantization we have chosen the jet direction, i.e., the $x$ axis, as the longitudinal direction. Consequently, the $x$-component of the spinor matrix $ \vec S=\vec \sigma/2$ contains the conventionally defined Pauli-$\sigma^3$ matrix 
\begin{equation}
    S^x=\frac{1}{2}\sigma^3 = \frac{1}{2}\begin{pmatrix}
    1 & 0 \\
    0 & -1
\end{pmatrix}\,.
\end{equation}
We identify the quark's magnetic momentum as $\vec\mu_q=  g_s g_q\vec S $, where the gyromagnetic ratio is $g_q\equiv g/p^+$ and the Land\'{e} g-factor $g_s=2$. This term is actually a reduced form of $-\vec\mu_q\cdot \mathcal{ \vec B}$, under the assumption that the longitudinal momentum transfer from the external field $\mathcal A^\mu$ is negligible compared to the $p^+$ momentum component of the jet. This implies that spin flips of the quark state are forbidden.

As we have previously mentioned, we retain terms suppressed as $1/p^+$ in \cref{eq:1stQuantHamiltonian} as they can contribute to the derivation of the quantum operators associated to the physical observables even in the limit $p^+ \to \infty$. After the relevant operators have been derived, we will take the eikonal limit on the Hamiltonian and retain only the terms which are not $p^+$ suppressed
\begin{equation}    \label{eq:EikonalHamiltonian}
   \lim_{p^+\to \infty} P^- = 2g \mathcal{A}_+\, .
\end{equation}
We will use the Hamiltonian in \cref{eq:EikonalHamiltonian} to compute the quantum evolution of the quark state, and the full Hamiltonian in \cref{eq:1stQuantHamiltonian} only when deriving quantum operators in \cref{sec:JetObservables}. Therefore, only the $\mathcal{A}_+$ component of the background fields enters the evolution Hamiltonian of the system, while the transverse $\vec{\mathcal{A}}_\perp$ components enter the definition of the kinetic momentum of the particle, as in \cref{eq:KineticMomentum}. The Zeeman term in the Hamiltonian does not play any role at leading eikonal order, nor does the $\mathcal{A}_-$ component of the background fields.

\subsubsection{Basis representation}
\label{subsec:basisrepr}

We employ a basis representation to construct the Hilbert space of the system and to perform the time evolution. We choose the eigenstates of the kinetic energy part of the free Hamiltonian as the basis states $\ket{\beta}$, so that the basis retains the same symmetries as the free theory
\begin{align}
    \frac{m^2 + \vec{p}_\perp^2}{p^+}\ket{\beta}=P^-_\beta\ket{\beta}\,.
\end{align}
Each basis state is labeled by five quantum numbers, the longitudinal momentum $p^+$, the transverse momentum $\vec p_\perp$, the light-front helicity $\lambda$, and the color index $c$. 

We discretize the transverse $(y,z)$ position space on a square lattice with extent $L_\perp$ and $N_\perp$ sites per dimension, such that the lattice spacing is $a_\perp=L_\perp/N_\perp$. In the $y$-direction this discretization matches that used for the Glasma fields, described in \cref{sec:GlasmaFields}. The associated momentum space is also discretized on a finite lattice by Fourier transforming the coordinate space lattice. Accordingly, the transverse momentum resolution is $d_p=2\pi/L_\perp$, acting as an effective infrared (IR) cutoff for the momentum transferred by the background fields, while the finite lattice spacing introduces an effective ultraviolet (UV) cutoff $\lambda_{UV}=\pi/a_\perp$. The transverse momentum in the basis space can therefore be written as
\begin{align}
    p_i = k_i d_p\,,\qquad i=y,z\;,
\end{align}
with the dimensionless quantum number $k_i=-N_\perp/2, -N_\perp/2+1,\ldots, N_\perp/2-1 $. Since the background field preserves the longitudinal momentum and the light-front helicity, we label the basis state as $\beta=\{ k^y, k^z,c\} $. 

The jet state can then be expanded as
\begin{equation} \label{eq:QuarkStateExpansion}
    \ket{\psi; x^+} = \sum_\beta c_\beta(x^+) \ket{\beta}\, ,
\end{equation}
where $c_\beta(x^+)\equiv\braket{\beta| \psi; x^+} $ are the basis coefficients at $x^+$. All the information about time evolution is now contained in the coefficients, which can be collected into a column vector $\mathbf{c}(x^+)$. This formulation greatly simplifies the computation, as the evolution reduces to matrix multiplications between the coefficient vector and the corresponding evolution matrix.

\subsubsection{Time evolution}

The evolution of the jet as a quantum state obeys the time-dependent Schr\"{o}dinger equation on the light front
\begin{align} \label{eq:SchrodingerEq}
    i \frac{\partial}{\partial x^+} \ket{\psi, x^+} = \frac{1}{2}P^-(x^+) \ket{\psi; x^+}\,.
\end{align}
The solution of the evolved state can be written as
\begin{align} \label{eq:SchrodingerSolution}
    \ket{\psi; x^+} = U( x^+; 0) \ket{\psi; 0}\,,
\end{align}
with the unitary evolution operator defined as
\begin{align} \label{eq:U}
    U( x^+_f; x^+_i) = \mathcal{T}_+ \exp\left[ -\frac{i}{2} \int_{x^+_i}^{x^+_f} \diff x^+ P^-(x^+) \right] \,.
\end{align}
Here, $\mathcal{T}_+  $ denotes the light-front time ordering.
In the basis representation, where the state is expanded as in \cref{eq:QuarkStateExpansion}, the solution takes the matrix form
\begin{equation} \label{eq:CoefficientEvolution}
    \textbf{c}(x^+) = \mathcal{T}_+ \exp\left[-\frac{i}{2}  \int_0^{x^+} \diff s \mathcal{M}(s) \right] \textbf{c}(0)\; ,
\end{equation}
where $\mathcal{M}_{\beta\beta^\prime}(x^+) \equiv \bra{\beta}P^-\ket{\beta^\prime}$ are the matrix elements of the light-front Hamiltonian.

Knowing the quantum state, an observable can be evaluated as the expectation value of the corresponding operator $\mathcal{O}$, 
\begin{align} \label{eq:Schrodinger_expval}
\begin{split}
    \braket{\mathcal{O}(x^+)}_\psi = &\braket{\psi; x^+|  \mathcal{O}(x^+) |\psi; x^+ } \\
    =  &\sum_{\beta \beta^\prime} c^*_\beta(x^+) c_{\beta^\prime} (x^+) \mathcal{O}_{\beta\beta^\prime}(x^+)\; .
    \end{split}
\end{align}
where $\mathcal{O}_{\beta\beta^\prime}(x^+) \equiv \bra{\beta}O\ket{\beta^\prime}$ denotes the matrix elements of the operator $\mathcal{O}$ in the chosen basis. One should not confuse the quantum expectation value of an operator, noted as $\braket{...}_\psi$ with the average over Glasma configurations $\braket{...}_{event}$ defined in \cref{eq:ConfigAvg}.

We formulate the time evolution in the Schr\"{o}dinger picture, noting that it can be expressed equivalently in the interaction or Heisenberg picture. The interaction picture is often preferred when studying the effect of an external field on the eigenstates of the free Hamiltonian, as in studies of a quark jet traversing a medium field~\cite{Li:2020uhl,Li:2025wzq}. In the eikonal limit relevant to our calculation, only the interaction part of the Hamiltonian survives, and thus the Schr\"{o}dinger and interaction pictures coincide. This is the picture we adopt for the quantum calculation. The Heisenberg picture will also be employed in later sections, as it facilitates comparison with the classical treatment in our companion work~\cite{Avramescu:2026fgv}, where it is likewise adopted. In particular, the derivation of the kinetic momentum operator becomes more transparent in the Heisenberg picture, since the time dependence is carried by the position operator.

The operators in the Heisenberg picture are related to those in the Schr\"odinger picture by
\begin{align}
    \mathcal O_H (x^+) = U^\dagger(x^+;0) \mathcal \,O (x^+)\, U(x^+;0)\;,
\end{align}
and the states are time-independent
\begin{align}
    \ket{\psi}_H \equiv\ket{\psi;0}= U^\dagger(x^+;0) \ket{\psi;x^+}\,.
\end{align}
The time evolution of the system is encoded at the operator level through Heisenberg's equation of motion,
\begin{align}\label{eq:HeisenbergOperatorEvolution}
    \frac{\diff \mathcal{O}_H}{\diff x^+} = i \left[\frac{1}{2}P_H^-, \mathcal{O}_H\right] + \frac{\partial \mathcal{O}_H}{\partial x^+}\, ,
\end{align}
and the evaluation of observables is equivalent to that in the Schr\"{o}dinger picture
\begin{align} \label{eq:Heisenberg_expval}
    \langle\mathcal{O}(x^+)\rangle_\psi = \,_H\langle \psi|  \mathcal{O}_H(x^+) |\psi \rangle_H \,.
\end{align}

\section{Jet observables}\label{sec:JetObservables}

In this section, we derive the quantum operators associated with the jet observables of interest. Specifically, we focus on the momentum broadening induced by the background field and the subsequent jet quenching parameter. We also introduce the color transition rate to quantify the effect of the medium on the color charge of the quark jet.

\subsection{Momentum broadening}\label{sec:broadening_calc}

It is well established, both in classical and quantum mechanics, that for a particle propagating through an external medium, the momentum conjugated to the position operator---the canonical momentum $\vec{p}_\perp$---does not generally coincide with the physical kinetic momentum, which we denote by $\vec{p}_\perp^{\, kin} $. In nonrelativistic quantum mechanics, the kinetic momentum of a particle is defined as mass times velocity, where the velocity is given by the time derivative of the position operator.
In light-front dynamics, the longitudinal momentum $p^+$ plays the role of particle mass in the nonrelativistic theory, as it is the central charge of the transverse Galilei subgroup. The transverse kinetic momentum is therefore defined as 
\begin{align}\label{eq:pkin_def}
    \vec{p}_\perp^{\, kin} = p^+ \frac{\diff \vec{x}_\perp}{\diff x^+}\;.
\end{align}

The time derivative of the coordinate operator is readily obtained in the Heisenberg picture using the evolution equation given in \cref{eq:HeisenbergOperatorEvolution} and the Hamiltonian in the quantum many-body representation given in \cref{eq:1stQuantHamiltonian}
\begin{align}
    \frac{d\vec{x}_{\perp, H}}{dx^+} &= i\left[\frac{1}{2}P_H^-, \vec{x}_{\perp, H}\right] \notag \\
    &= U^\dagger(x^+; 0) \frac{\vec{p}_\perp - g \vec{\mathcal{A}}_\perp}{p^+} U(x^+; 0) \, ,
\end{align}
where we have applied the canonical commutation relations $[p_i, x_i] = -i$. The quantum operator of transverse kinetic momentum is therefore
\begin{equation}\label{eq:KineticMomentumOperator}
    \vec{p}^{\, kin}_\perp = \vec{p}_\perp - g \vec{\mathcal{A}}_\perp\;.
\end{equation}
The kinetic momentum is the physical, gauge-invariant quantity, whereas the canonical momentum $\vec{p}_\perp$ includes a gauge-dependent contribution from the external field. We emphasize that when referring to ``momentum space", particularly in the context of basis representation, we mean the canonical momentum. 

In deriving the kinetic momentum operator, we employ the full light-front Hamiltonian in \cref{eq:1stQuantHamiltonian}. This is necessary because the operator is proportional to $p^+ P^-$, so terms in the Hamiltonian that are formally suppressed by $1/p^+$ still contribute to the observable, even in the eikonal limit $p^+ \to \infty$. This treatment is therefore consistent with—indeed required for—the use of the eikonal Hamiltonian in \cref{eq:EikonalHamiltonian} to describe the time evolution.

To quantify the transverse momentum broadening over a finite evolution time, we introduce the operator
\begin{align} \label{eq:DirectBroadOperatorDefiniiton}
   (\delta p^{kin}_{i, H}(x^+))^2= (p^{kin}_{i, H}(x^+) -  p^{kin}_{i, H}(0))^2 \; ,
\end{align}
where $i \in (y,z)$ and we do not sum over $i$, to study the anisotropies in the momentum broadening between the two directions transverse to the jet.
This definition removes the contribution from a nonzero initial kinetic momentum, which can arise when the jet is initialized as a wave packet in transverse momentum space or when the background field is nonzero initially. 

In the above equation, we express the operator in the Heisenberg picture to avoid ambiguities associated with time dependence. In contrast, in the Schr\"{o}dinger picture both the quantum state and the Glasma field $\vec{\mathcal{A}}_\perp $, and consequently $ \vec{p}^{\,kin}_\perp$, carry explicit time dependence.
The operator in the Schr\"{o}dinger picture is obtained via the standard unitary transformation
\begin{equation}
    (\delta p^{kin}_{i}(x^+))^2= U(x^+;0)\,(\delta p^{kin}_{i, H}(x^+))^2\, U^\dagger(x^+;0)\;.
\end{equation}

We evaluate the corresponding expectation value using two complementary approaches, detailed in the following subsections: a direct quantum calculation based on the operator definition of $p^{kin}_{i}$ given in \cref{eq:KineticMomentumOperator}, and an alternative formulation in terms of Lorentz force quantum operator.

\subsubsection{Direct quantum calculation} \label{sec:direct}

The expectation value of the momentum broadening operator defined in \cref{eq:DirectBroadOperatorDefiniiton} can be evaluated equivalently in either the Heisenberg or Sch\"{o}dinger picture, 
\begin{align}
\begin{split}
    \braket{(\delta p^{kin}_{i} (x^+))^2 }_\psi
    =&_H\braket{\psi|(\delta p^{kin}_{i}(x^+) )^2_H|\psi }_H\\
    =&\braket{\psi;x^+|(\delta p^{kin}_{i}(x^+) )^2 |\psi;x^+ }\;.
\end{split}
\end{align}
We adopt the latter formulation here. Expanding the square in \cref{eq:DirectBroadOperatorDefiniiton}, three terms contribute 
\begin{multline} \label{eq:DirectBroadOperatorExpval}
\braket{(\delta p^{kin}_{i} (x^+) )^2 }_\psi = \langle (p^{kin}_{i}(x^+))^2 \rangle_\psi  + \langle (p^{kin}_{i}(0))^2_\psi \rangle \\
     - 2 \Re \langle p^{kin}_{i}(x^+) p^{kin}_{i}(0)  \rangle_\psi  \, ,
\end{multline}
where $\Re$ denotes the real part. The two squared terms are straightforward to evaluate,
\begin{align}
\begin{split}
    \braket{(p^{kin}_{i}(x^+))^2 }_\psi  
    & = \braket{\psi;x^+|(p^{kin}_{i}(x^+))^2|\psi;x^+ }\\
    & = \braket{ \psi;x^+|\left( p_{i}-g\mathcal A_i(x^+)\right)^2|\psi;x^+ }
    \;,
\end{split}
\end{align}
and 
\begin{align}
\begin{split}
    \braket{(p^{kin}_{i}(0))^2 }_\psi  
    & = \braket{ \psi;0|\left( p_{i}-g\mathcal A_i(0)\right)^2|\psi;0 }
    \;.
\end{split}
\end{align}
The cross term requires additional manipulations and can be written as 
\begin{align}
\begin{split}
\langle p^{kin}_{i}& (x^+)  p^{kin}_{i}(0)  \rangle_\psi \\
  & = \braket{ \psi; 0| U^\dagger(x^+; 0) p^{kin}_i(x^+) 
    U(x^+; 0) p^{kin}_i(0) | \psi; 0}\\
    & = \braket{ \psi; x^+| p^{kin}_i(x^+) 
| \tilde{\psi}_i; x^+}\\
 & = \braket{ \psi; x^+| p_{i}-g\mathcal A_i(x^+)
| \tilde{\psi}_i; x^+}\;,
\end{split}
\end{align}
where we have defined an auxiliary state 
\begin{align}
    \ket{ \tilde{\psi}_i; 0}\equiv p^{kin}_i(0) \ket{ \psi; 0}\,,
    \quad
    \ket{ \tilde{\psi}_i; x^+}=U(x^+;0)\ket{ \tilde{\psi}_i; 0}\;.
\end{align}
In practice, the auxiliary state $\ket{ \tilde{\psi}_i; 0}$ is prepared alongside the original state $\ket{ \psi; 0}$ and evolved simultaneously under the same Hamiltonian, so that the momentum broadening observable can be evaluated at arbitrary light-front time $x^+$.

Let us now consider the particular case where the background fields are the Glasma fields. The time-evolution operator \cref{eq:U} reduces to the path-ordered exponentiation of the longitudinal field component
\begin{align} \label{eq:TimeEvolutionOperator}
    \begin{split}
        U_+(x^+; 0) =~  & \mathcal{T}_+ \exp\left[ -ig \int_0^{x^+} \diff s^+ \mathcal{A}_+ (s^+, y, z) \right] \;,
    \end{split}
\end{align}
where the Glasma field $\mathcal{A}_+$ are evaluated according to \cref{eq:ApproxJetTransformation}, which incorporates the mid-rapidity expansion. The transverse field components $\mathcal{A}_y$ and $\mathcal{A}_z$ entering the kinetic momentum operator are likewise given by \cref{eq:ApproxJetTransformation}.

As discussed in \cref{sec:GlasmaFields}, the Glasma fields fluctuate across events due to the random initial color charge configurations sampled from the MV model \cref{eq:MVmodel}. Accordingly, the momentum broadening must be averaged over a large ensemble of Glasma events to remove the dependence on a particular color charge realization. We adopt the combined notation
\begin{align}\label{eq:combined_ave}
    \braket{\cdots} \equiv \braket{\braket{\cdots}_\psi}_{event}
\end{align}
to indicate both the quantum expectation value defined in \cref{eq:Schrodinger_expval} and the average over Glasma color charges configurations in \cref{eq:ConfigAvg}, and use this notation throughout the remainder of this manuscript.

\subsubsection{Lorentz force calculation} \label{sec:quantLorentz}

An alternative approach to evaluating the momentum broadening operator is to compute the change rate of the transverse kinetic momentum and then integrate this operator over the evolution time 
\begin{align}
    \braket{(\delta  p^{kin}_i (x^+))^2}_\psi =& \int_0^{x^+} \diff s^+ \int_0^{x^+} \diff \bar{s}^+ \notag \\
    &\times \prescript{}{H}{\bra{\psi}} \frac{\diff {p}^{\, kin}_{i, H}(s^+)}{\diff s^+}
   \frac{\diff {p}^{\, kin}_{i, H}(\bar{s}^+)}{\diff \bar{s}^+}\ket{\psi}_H\;.
\end{align}
which provides a direct correspondence to the classical calculation based on Wong's equation in \cref{sec:ClassicPropagation} and Ref.~\cite{Ipp:2020mjc}.

In the Heisenberg picture, the light-front time derivative of the kinetic momentum operator gives the Lorentz force experienced by the jet as
\begin{equation}\label{eq:QuantumLorentzForce}
    \frac{\diff p^{kin}_{i, H}}{\diff x^+} = \frac{g}{2} f_{i, H} (x^+)\,.
\end{equation}
Applying the Heisenberg equation of motion in \cref{eq:HeisenbergOperatorEvolution}
\begin{equation} \label{eq:pkinHeisenbergEquation}
    \frac{\diff p^{kin}_{i, H}}{\diff x^+}=i\left[\frac{1}{2}P_H^-, p^{kin}_{i, H}\right] 
    +  \frac{\partial p^{kin}_{i,H}}{\partial x^+}\,,
\end{equation}
yields the Lorentz force operator in the Schr\"{o}dinger picture (see \cref{app:Lorentz})
\begin{align} \label{eq:LorentzForceOperator}
    f_i (x^+) = 2\, \mathcal{F}_{i+} (x^+)\,,
\end{align}
where $\mathcal{F}_{\mu\nu}$ is the field strength tensor of the classical background fields. This expression is in close analogy with the classical non-abelian Lorentz force for an eikonal particle propagating through the Glasma, as in \cref{eq:f_classical} and Refs.~\cite{Ipp:2020mjc,Ipp:2020nfu,Avramescu:2023qvv,Avramescu:2026fgv}. Note that the terms suppressed as $1/p^+$ in the Hamiltonian \cref{eq:1stQuantHamiltonian} do not contribute here, since there are no additional $p^+$ factors in the operator definition and the eikonal Hamiltonian in \cref{eq:EikonalHamiltonian} suffices. In the Heisenberg picture, the Lorentz force operator becomes
\begin{equation} \label{eq:HeisenbergLorentzForce}
    f_{i, H} (x^+)=U^\dagger(x^+; 0)\,f_i(x^+)\,U(x^+; 0)\,.
\end{equation}

Substituting \cref{eq:QuantumLorentzForce} into the double-time integral, the kinetic momentum broadening reads 
\begin{align} \label{eq:LFTimeKineticBroadening}
    \langle(\delta  p^{kin}_i(x^+))^2\rangle_{\psi} & = \frac{g^2}{4} \int_0^{x^+} \diff s^+ \int_0^{x^+} \diff \bar{s}^+ \notag \\
    & \times \bra{\psi; 0} f_{i,H}(s^+) f_{i,H}(\bar{s}^+) \ket{\psi; 0}\;.
\end{align}
This expression is manifestly gauge invariant, with the evolution operator $U(x^+;0)$ playing the same role as the classical color rotation operator $\mathcal{U}(\tau;0)$ in \cref{eq:PartWilsonLine}: both perform the parallel transport of the Lorentz force along the jet trajectory.

We emphasize that an important distinction between \cref{eq:LFTimeKineticBroadening} and its classical counterpart in \cref{eq:KinMomBroadJet} lies in the treatment of color degrees of freedom.
In the classical approach, one solves the equations of motion for the color charge, average over color charge configurations according to \cref{eq:defintq} and the two-point function from \cref{eq:intqaqb}, and normalizes to obtain an overall Casimir factor. By contrast, in the quantum calculation, the color degrees of freedom are encoded directly in the quark state defined in \cref{eq:QuarkStateExpansion}, and the quantum expectation value is subsequently averaged over the initial quark color. The calculation is therefore fully quantum from the perspective of the jet, with the only classical approximation entering through the gluon background field, where it is justified by the large gluon occupation numbers. Furthermore, taking the classical limit of \cref{eq:QuantumLorentzForce} as discussed in \cite{Avramescu:2026fgv} recovers the classical momentum broadening result, demonstrating that the classical–quantum correspondence is consistently realized within the light-front quantization framework.

So far, the derivation in this section has been carried out for a generic classical background field. We can now specify the result for the background Glasma fields. Substituting the rapidity expanded fields from \cref{eq:ApproxJetTransformation} into \cref{eq:LorentzForceOperator}, the relevant components of the Lorentz force operator are
\begin{equation}\label{eq:MilneFieldStrength}
    \begin{split}
        f_y(x^+) & =- \left( \partial_\tau \mathcal{A}_y + \mathcal{F}_{xy} + \frac{4z}{(x^+)^2} \mathcal{D}_y \mathcal{A}_\eta \right) \, ,  \\
        f_z (x^+) & =- \frac{2}{x^+} \left(\partial_\tau \mathcal{A}_\eta +\mathcal{D}_x\mathcal{A}_\eta\right)\; .
    \end{split}
\end{equation}
where the gauge fields $\mathcal{A}_\mu(x^+)$ have a coordinate dependence according to \cref{eq:notamuxplus} and are expressed as in \cref{eq:ApproxJetTransformation}. Since the Lorentz force calculation is carried out entirely in coordinate space, no measurement of canonical momentum is required. We can therefore initialize the wavefunction as a delta function in coordinate space at $y=z=0$, which eliminates the last term in $f_y$. Setting $x^+ = 2\tau$, the electric and magnetic field components entering the jet evolution are the ones given by \cref{Eq:EBFieldsGlasmaClass} and the Lorentz force quantum operator takes the form 
\begin{align} \label{eq:JetLorentzForce}
    f^y\vert_{y=z=0}=\mathcal{E}_y-\mathcal{B}_z,\quad f^z\vert_{y=z=0}=\mathcal{E}_z+\mathcal{B}_y\;,
\end{align}
which also holds for general $y$. The time evolution operator also reduces to 
\begin{align} \label{eq:LoretnzColorRotation}
    U(x^+; 0) = \mathcal{T}_+ \exp\left[ -i\frac{g}{2} \int_0^{x^+} \diff s^+ \left. \mathcal{A}_x \left(s^+ \right) \right|_{y=z=0} \right] \, ,
\end{align}
with the fields evaluated at $\tau = x = s^+/2$. The Lorentz force components in \cref{eq:JetLorentzForce} together with the color rotation in \cref{eq:LoretnzColorRotation} are substituted into \cref{eq:LFTimeKineticBroadening} to obtain the kinetic momentum broadening for a single Glasma event, which is then averaged over Glasma configurations as in \cref{eq:ConfigAvg}.

\subsubsection{Delocalization effects: Abelian toy-model} \label{sec:ToyModel}

In this section, we explore how the finite spatial extent of the jet in transverse space affects momentum broadening. We start from the Lorentz force components in \cref{eq:MilneFieldStrength}, but instead of treating the jet as a perfectly localized delta function, we consider a Gaussian wave packet
\begin{align} \label{eq:GaussianWavefunction}
\begin{split}
    \ket{\psi} =& \frac{1}{(2\pi\sigma_y^2)^{1/4}} \frac{1}{(2\pi\sigma_z^2)^{1/4}} \int \diff y \diff z\, 
    \\
    &\times \exp\left[-\frac{y^2}{4\sigma_y^2}\right]
    \exp\left[-\frac{z^2}{4\sigma_z^2}\right]
    \ket{y, z} \; ,
\end{split}
\end{align}
where the color and helicity indices are absent as they are not relevant for this calculation. Let us now consider, for simplicity, that the background field is abelian so the color rotation operators cancel. Using that the field strength tensor Heisenberg operator is local in coordinate space we can write $\bra{y,z} f_{i,H}(s^+) f_{i,H}(\bar{s}^+)\ket{y', z'} = f_{i,H}(s^+) f_{i,H}(\bar{s}^+) \delta(y-y^\prime)\delta(z-z^\prime)$. By substituting the wavepacket in \cref{eq:GaussianWavefunction} into \cref{eq:LFTimeKineticBroadening} we obtain the kinetic momentum broadening for a single Glasma event as
\begin{align}\label{eq:pkin_ff}
    \braket{(\delta p_i^{kin})^2}_\psi & = \frac{g^2}{4} \frac{1}{(2\pi\sigma_y^2)^{1/2}} \frac{1}{(2\pi\sigma_z^2)^{1/2}} \notag \\
    & \times \int \diff y \diff z \exp\left[-\frac{y^2}{4\sigma_y^2}\right]
    \exp\left[-\frac{z^2}{4\sigma_z^2}\right] \notag \\
    & \times \int_0^{x^+} \diff s^+ \int_0^{x^+} \diff \bar{s}^+ \left[f_i(s^+) f_i(\bar{s}^+)\right] \, ,
\end{align}
where $i=y,z$, and the Lorentz force is given by \cref{eq:MilneFieldStrength}, valid for small deviations from mid-rapidity.

To perform the integration over the $(y,z)$ coordinates we use that, after averaging over Glasma configurations as in \cref{eq:ConfigAvg}, any point in the $y$ direction becomes equivalent, so $\braket{F[\mathcal{A}(y)]}_{event} = \braket{F[\mathcal{A}(y=0)]}_{event}$, with $F[\mathcal{A}]$ an arbitrary local function of the background fields $\mathcal{A}$. The $y$ dependence in the Lorentz force can therefore be dropped and the $y$ integral is trivial. 
The $z$ integration is less trivial: as seen in \cref{eq:ApproxJetTransformation}, $\mathcal{A}_+$ carries an explicit dependence on $z$ arising from the transformation between Milne and light-cone coordinates, which is not removed by the configuration average. The relevant Gaussian integrals over $z$  are
\begin{subequations}
    \begin{align}
        \int_{-\infty}^\infty \diff z\,  \frac{z}{(2\pi \sigma_z^2)^{1/2}} \exp\left[-\frac{z^2}{2\sigma_z^2}\right] & = 0 \, , \\
        \int_{-\infty}^\infty \diff z\,  \frac{z^2}{(2\pi \sigma_z^2)^{1/2}} \exp\left[-\frac{z^2}{2\sigma_z^2}\right] & = \sigma_z^2 \;.
    \end{align}
\end{subequations}
The configuration-averaged momentum broadening then reads
\begin{subequations} \label{eq:FiniteSizeEffects}
    \begin{align} 
        \braket{(\delta p_y^{kin})^2}  & = \frac{g^2}{4} \int_0^{x^+} \diff s^+ \int_0^{x^+} \diff \bar{s}^+ \left\langle \left[f_y(s^+) f_y(\bar{s}^+)\right]_{y=z=0} \right. \notag \\
        & \hspace{-20 pt}  \left. + \sigma_z^2\left[ \frac{4}{(s^+)^2} \partial_y \mathcal{A}_\eta\left(s^+ \right) \frac{4}{(\bar{s}^+)^2} \partial_y \mathcal{A}_\eta\left(\bar{s}^+ \right)\right]_{y=0} \right\rangle_{event} , \label{eq:FiniteSizeTransverse} \\
        \braket{(\delta p_z^{kin})^2}  & = \frac{g^2}{4} \int_0^{x^+} \diff s^+  \int_0^{x^+} \diff \bar{s}^+ \notag \\
        & \hspace{40 pt} \times \left\langle \left[f_z(s^+) f_z(\bar{s}^+)\right]_{y=z=0} \right\rangle_{event} \label{eq:FiniteSizeLongitudinal}\, ,
    \end{align}
\end{subequations}
where $f_z(x^+)|_{y=z=0}$ is the Lorentz force in \cref{eq:JetLorentzForce} and the $\mathcal{A}_\eta$ components are evaluated at $\tau=x = x^+/2$ as explained in \cref{sec:Glasma_jetLC}. Recall that $\braket{\cdots}$ indicates both the quantum expectation value and the average over Glasma configurations, as defined in \cref{eq:combined_ave}.

Several observations follow from these results. First, the kinetic momentum broadening is independent of $\sigma_y$, the jet width in the $y$ direction, as expected from the translational invariance of the system after configuration averaging. Second, broadening in the $z$ direction is also independent of $\sigma_z$: since $\mathcal{A}_+$ depends only linearly on $z$, this dependence is removed when taking the derivative $\partial_z \mathcal{A}_+$ to form $f_z$ as given in  \cref{eq:LorentzForceOperator}. Third, and most notably, \cref{eq:FiniteSizeTransverse} contains a term proportional to $\sigma_z^2$ that is absent for a perfectly localized jet as in \cref{Eq:EBFieldsGlasmaClass}. This indicates that $\langle(\delta p_y)^2\rangle$ is sensible to the jet width in $z$ direction, deviating from the localized particle results. At early times $\mathcal{A}_\eta \propto \tau^2$, so the ratio $\mathcal{A}_\eta/\tau^2$ remains finite as $\tau \to 0$, and this correction can be safely neglected for strongly collimated jets even at early times.

A quantitative study of these delocalization effects is, however, beyond the reach of the present approximation scheme. The $z$-expanded fields in \cref{eq:ApproxJetTransformation} are valid only when $\sigma_z\, (g^2\mu) \ll 1$, i.e., when the transverse extent of the jet is small enough that $\sigma_z\,( g\mathcal{A}_\mu)$ can be considered as a small correction. The delocalization correction becomes significant precisely in the regime where this condition is violated. One might attempt to circumvent this limitation by retaining the exact coordinate transformation between Milne and Minkowski coordinates. However, as discussed in \cref{sec:GlasmaFields}, the finite spatial extent of the Glasma fields restricts the jet widths that can be simulated without sampling space-time points outside the causal domain, so in practice this approach does not extend our ability to study wide jets. We expect, however, the qualitative behavior to hold in the physical situation. 

It is important to note that the observed sensitivity to $\sigma_z$ is not inherently quantum, but rather arises from probing space-time points away from $z=0$, where the fields differ from their values at the origin. A Gaussian distribution of classical probe particles would yield the same correction as in \cref{eq:FiniteSizeTransverse}. Nevertheless, the effect becomes particularly relevant in the quantum framework, where particles are naturally described as distributions in both coordinate and momentum space.

\subsection{The jet quenching parameter}\label{sec:qhat_calc}

The jet quenching parameter $\hat{q}$ is a central quantity that provides information about the medium probed by the jet. It is typically extracted from experimental data, see e.g. \cite{Apolinario:2022vzg} for a recent review of different phenomenological analyses, and serves as an input parameter for calculations in the QGP phase. Here, we compute the transport coefficient $\hat q$ in the Glasma as the amount of squared momentum transferred from the fields to the jet per unit time
\begin{equation} \label{eq:qhat}
    \hat{q} = 2\, \frac{\diff \langle (\delta \vec{p}^{\, kin}_\perp)^2 \rangle}{\diff x^+}\, .
\end{equation}
The factor of $2$ is chosen in order to match the usual definition in literature, which uses a derivative with respect to $t$. We are interested in how this parameter evolves in time, as it serves as a probe of the evolution and strength of the background fields. On dimensional grounds, one expects the jet quenching parameter to scale as $\hat{q} \propto Q_s^3$, where $Q_s$ is the saturation momentum, that sets the characteristic scale of the Glasma fields. The saturation scale is related to the color charge density $g^2\mu$ and the IR regulator $m_g$ through the Wilson line correlator \cite{Lappi:2007ku}. The presence of $m_g$ as an additional dimensionful parameter can break the pure $Q_s^3$ scaling, though we expect this behavior to hold approximately.

The formalism of jet quenching in the QGP assumes that the parton experiences kicks from scattering centers uncorrelated in the direction of the propagation. Under this condition, there is a one-to-one correspondence between broadening and energy loss by soft gluon radiation in the high energy limit, as both are proportional to $\hat q$. Parametrically
\cite{Baier:1996kr}
\begin{equation} \label{eq:EnergyLossEstimate}
    \Delta E \sim \alpha_s \hat{q}\, L^2\;,
\end{equation}
where $\alpha_s$ is the strong coupling constant and $L$ the length of the particle path inside the medium
\footnote{
Note that the path length traveled by the particle inside the medium is related to the light-front propagation time by $L=x^+/2$, following our convention for light-front coordinates, $x^+=t+x$.
}.
The assumption of independent transverse momentum kicks breaks down in the Glasma, due to the coherence and the correlation structure of the color fields. In the absence of a full computation of medium-induced gluon radiation using Glasma fields -- see, however the recent work \cite{Barata:2024xwy} -- that would test its validity, \cref{eq:EnergyLossEstimate} should be regarded as an educated estimate of the possible magnitude of pre-equilibrium energy loss relative to that in the QGP phase.

\subsection{Color rotation}\label{sec:rotation_calc}

The time evolution of the quantum state is governed by the path-ordered exponential in \cref{eq:SchrodingerSolution} with the eikonal Hamiltonian in \cref{eq:EikonalHamiltonian}. This exponential represents a color rotation of the quark state: a quark initially in a definite color state is rotated into a superposition of color states through its interaction with the Glasma field. The time scale over which this color homogenization occurs depends on the properties of the field, in particular on the saturation scale $Q_s$. It is subsequently of interest to study this color rotation as a means of extracting information about the properties of the Glasma.

The probability of finding the quark in a given color state $c$ at light-front time $x^+$ is
\begin{align} \label{eq:ColorEvolution}
    P_c(x^+) = |\langle \psi; x^+|c\rangle |^2\;,
\end{align}
with the normalization condition 
\begin{align}
    \sum_{c=1}^{N_c} P_c(x^+)=1\;,
\end{align}
satisfied at all times. As for the momentum broadening, the color rotation is averaged over a large ensemble of Glasma configurations to eliminate dependence on the initial classical color charge fluctuations.

It is important to note that the color of a quark state is not a physically observable quantity. Indeed, \cref{eq:ColorEvolution} is not gauge invariant, and consequently the color homogenization timescale depends on the choice of gauge for the background fields. It is expected that, in gauges where the magnitude of the potential is large, the color mixing proceeds more rapidly, while in gauges where the potential is small---such as Coulomb gauge \cite{Avramescu:2026fgv}---homogenization is expected to take longer.

\section{Results and discussion}\label{sec:results}

Let us now move to presenting our  results for momentum broadening, the jet quenching parameter, and color rotation effects obtained from simulations of jet evolution in the Glasma. The Glasma fields and the jet state are discretized on the same transverse lattice, with parameters chosen such that $g^2\mu L_\perp = 100$ and $N_\perp = 1024$.  This ensures that the lattice spacing $a_\perp = L_\perp/N_\perp$ is sufficient to resolve the Glasma flux tubes of characteristic size $1/Q_s \sim 1/(g^2\mu)$. For standard values $g^2\mu \approx 1$--$2.5~\GeV$, these parameters correspond to system sizes similar to the area of a large nucleus, $L_\perp\approx 8$--$20 \fm$. We further require that the quantum wave packet is well contained within the lattice, meaning that the probability of occupying momentum states above the ultraviolet cutoff $\lambda_\mathrm{UV} = \pi/a_\perp$ is negligible. For every observable and every parameter configuration we run $100$ Glasma events. We extract the expectation value of the observable for each of them and then average over all the events to obtain the configuration averaged magnitude as given by \cref{eq:ConfigAvg}. For the quantum results, $\braket{\ldots}$ denotes the combined quantum expectation value and Glasma configuration average, in the notation introduced in \cref{eq:combined_ave}. When performing the classical simulations to compare with the quantum results we only run $10$ Glasma events, but we do not only average over events and the classical color charge (as defined in \cref{eq:defintq}) but also over all the possible starting points of the jet in the $(x,y)$ plane, giving an effectively much larger number of configurations. In this case $\braket{\ldots}\to\braket{\braket{\braket{\ldots}_Q}_{\vec{r}_0}}_{event}$ denotes the average over the quark classical color charge, as defined in 
\cref{eq:defintq}, followed by the event average over Glasma configurations, which includes both the event average and the average over the jet starting point. To improve numerical accuracy, we fix the residual gauge freedom by imposing the transverse Coulomb gauge condition $\partial_{\underline{i}} \mathcal{A}_{\underline{i}} = 0$ \cite{Avramescu:2026fgv} at the initial time $\tau$, in addition to the Fock-Schwinger gauge condition $\mathcal{A}_\tau = 0$ used in the formulation of the Glasma fields.

\subsection{Momentum broadening }\label{sec:broadening_results}

Let us now extract the jet momentum broadening in the $y$ and $z$ directions both via the direct calculation in \cref{sec:direct} and the Lorentz force calculation in \cref{sec:quantLorentz}. We also follow the classical evolution of the probe as described in \cref{sec:ClassicPropagation}. Here we will plot our results in physical units assuming $g^2\mu = 2~\GeV$, which corresponds to a lattice transverse length $L_\perp = 50~\GeV^{-1} \approx 10~\fm$. The IR regulator from \cref{eq:Poisson} is set to $m_g = 0.2~\GeV$, ensuring $\lambda_\mathrm{IR} = 2\pi/L_\perp < m_g$ so that the physics is controlled by the Glasma model parameters rather than the lattice spacing or volume.

\begin{figure}[t]
\centering
\includegraphics[width=\columnwidth]{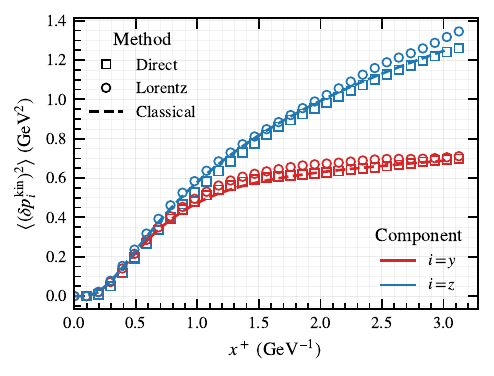}
\caption{Momentum broadening of a quantum particle traversing the Glasma fields. Squares (circles) represent the results obtained via the direct (quantum Lorentz force) calculation according to \cref{eq:DirectBroadOperatorExpval} (\cref{eq:LFTimeKineticBroadening}), while the dashed line corresponds to the classical probe particle simulation, according to \cref{eq:KinMomBroadJet}.  Results are shown for both the $y$ (red) and $z$ (blue) directions.
}
\label{fig:BroadeningComparison}
\end{figure}

We first consider a jet well localized around $y=z=0$, with $\sigma_y^2 = \sigma_z^2 = 0.005\, \GeV^{-2}$. In \cref{fig:BroadeningComparison} we show the results for the direct and the Lorentz force calculations, given by \cref{eq:DirectBroadOperatorExpval} and \cref{eq:LFTimeKineticBroadening} respectively, together with the classical results, obtained by solving the Wong's equations as in \cref{eq:KinMomBroadJet}. We find a good agreement between the two quantum approaches, an indicator of the robustness of our calculation.
We also find the quantum results to match the classical calculation. This is expected within our Fock space truncation to the $\ket{q}$ sector; by restricting to single-quark states we neglect gluon emissions and therefore quantum corrections to the quark propagation, so that our quantum formalism naturally reproduces the classical predictions. The result then serves as a consistency check for the formalism developed here, which will be extended in future works to include the effects of gluon radiation. Small deviations from a perfect agreement are interpreted as numerical uncertainties. Imposing the Coulomb gauge condition on the background field \cite{Avramescu:2026fgv} is essential here, without this additional gauge fixing these deviations would be much larger.

Moreover, \cref{fig:BroadeningComparison} reveals a clear asymmetry between the momentum broadening in the $y$ and $z$ directions, with broadening being systematically larger along the collision axis ($z$). This directional asymmetry is consistent with the findings of previous works~\cite{Ipp:2020mjc,Avramescu:2023qvv}, which report a similar anisotropy for classical probe particles propagating in Glasma fields. Notably, Ref.~\cite{Avramescu:2023vld} finds that for eikonal quarks, the ratio $\langle (\delta p_z^{kin})^2\rangle/\langle (\delta p_y^{kin})^2\rangle$ starts at $1$ at $\tau=0$ and quickly becomes anisotropic. Such anisotropic momentum broadening, acquired during the pre-equilibrium phase, persists into the subsequent thermalized stage, where it combines with anisotropies generated by collective flow or energy density gradients in the QGP~\cite{Armesto:2004pt,Baier:2008js,Sadofyev:2021ohn,Hauksson:2021okc,Barata:2022krd,Barata:2023qds,Barata:2024bqp,Barata:2025wnp,Andres:2022ndd,Hauksson:2023tze,Kuzmin:2023hko,Altenburger:2025iqa}, making a clean separation of the two contributions challenging. Nevertheless, the anisotropy in momentum broadening remains one of the most promising quantities for identifying imprints of the Glasma phase in experimentally measured jet properties such as the  jet substructure. 

\begin{figure*}[!htb] \label{fig:WidthDependence}
\centering

    \subfigure[
    \,Varying $\sigma_y^2$, fixed $\sigma_z^2$
    ]{\includegraphics[width=0.95\columnwidth]{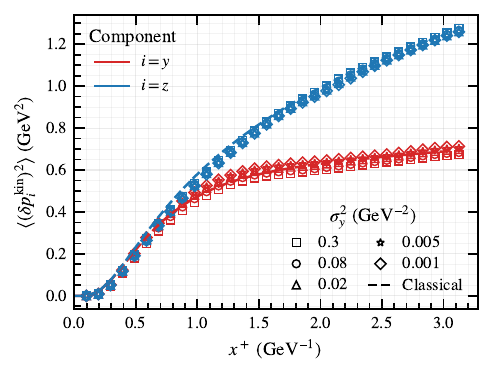} \label{fig:y_width_dependence}
    }
    \quad
    \subfigure[
    \,Varying $\sigma_z^2$, fixed $\sigma_y^2$]{\includegraphics[width=0.95\columnwidth]{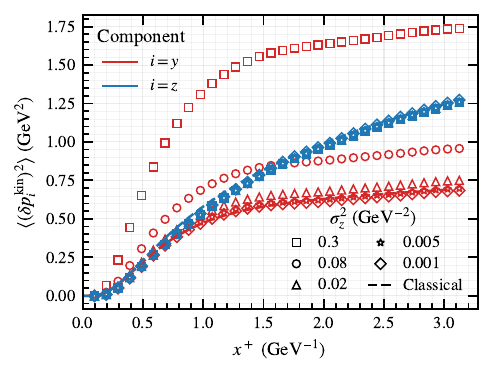} \label{fig:z_widthDependence}
    }
    
\caption{ 
Momentum broadening $\langle(\delta p_i^{\rm kin})^2\rangle$ as a function of the evolution time $x^+$ for varying the jet width (a) $\sigma_y^2$ and (b) $\sigma_z^2$. Red and blue curves show the $y$ and $z$ components, respectively. Different markers denote different width values. Dashed lines show the classical result, which corresponds to localized jets.
}
\label{fig:widthDependence}
\end{figure*}

Having validated our formalism against classical results for a completely localized jet, we now investigate how the transverse width of the jet wavefunction affects momentum broadening. Since both quantum approaches agree, we will present results obtained via direct evaluation of the kinetic momentum operator in the following, which is more in the usual spirit of evaluating observables as expectation values in a time dependent quantum state. \Cref{fig:widthDependence} shows the momentum broadening for jets of different transverse widths, also simulated using $N_{events}=100$ Glasma events as before, with the classical particle simulation for a localized jet serving as a reference.

In \cref{fig:y_width_dependence}, we fix $\sigma_z^2 = 0.005\, \GeV^{-2}$ and vary $\sigma_y^2$. Both $\langle(\delta p_y^{kin})^2\rangle$ and $\langle(\delta p_z^{kin})^2\rangle$ are insensitive to the width change, with all quantum results agreeing with the classical results for jets localized at $y=z=0$. We then fix $\sigma_y^2 = 0.005 \GeV^{-2}$ and vary $\sigma_z^2$ in \cref{fig:z_widthDependence}. While $\langle(\delta p_z^{kin})^2\rangle$ remains insensitive to the width change, $\langle(\delta p_y^{kin})^2\rangle$ grows significantly as the jet becomes more delocalized in the $z$ direction, developing large deviations from the localized classical result. This behavior is in perfect agreement with the analytical predictions derived in \cref{sec:ToyModel} using an Abelian toy model for the background fields. As follows from \cref{eq:FiniteSizeLongitudinal}, $\langle (\delta p_z^{kin})^2\rangle$ is independent of the transverse width of the jet in both $y$ and $z$ directions. In contrast, \cref{eq:FiniteSizeTransverse} contains a term proportional to $\sigma_z^2$, which causes $\langle(\delta p_y^{kin})^2\rangle$ to depend on the  width of the wave packet in the $z$ direction while remaining insensitive to $\sigma_y$. When $\sigma_z\,( g\mathcal{A}_\eta)$ is small, this term becomes negligible and the quantum results converge to the classical ones, recovered in the limit of a perfectly localized particle at $y=z=0$. 
However, as the jet gets wider the term containing $\sigma_z^2$ gains relevance, generating large deviations from the classical result.

It is worth emphasizing that the observed width dependence is not an intrinsic quantum effect, but rather a consequence of the finite spatial extent of the jet, as discussed in \cref{sec:ToyModel}.
An analogous effect would arise for a classical ensemble of probe particles distributed over a finite transverse area. Nevertheless, it appears naturally in the quantum calculation, where the particle inherently has a non-zero transverse size. Finally, we note that these results are obtained within a first order expansion of the fields in the $z$ coordinate, which breaks down for poorly localized jets. A more complete treatment would require retaining the exact transformation between Milne and light-cone coordinates and accounting for the finite transverse extent of the Glasma fields.

\subsection{The jet quenching parameter}

The jet quenching parameter $\hat{q}$ characterizes the momentum broadening induced on the jet per unit time, as defined in \cref{eq:qhat}. Here, we study how $\hat{q}$ scales with the physical saturation scale $Q_s$ and the IR regulator $m_g$, as discussed in \cref{sec:qhat_calc} and already observed in \cite{Ipp:2020mjc}. We also use the assumption, possibly very crude in the case of the Glasma fields,  of independent soft scatterings to turn these values of $\hat{q}$ into a rough estimate  of medium-induced energy loss in the Glasma phase. 

\subsubsection{Dependence on the IR regulator}

We consider three values of the charge density $g^2\mu = 1.33 \GeV$, $2 \GeV$ and $2.5 \GeV$ and two values of the IR regulator, $m_g = 0 \GeV$ and $0.2 \GeV$. The lattice transverse size $L_\perp$ is adjusted accordingly to keep $g^2\mu L_\perp = 100$ fixed, while $N_\perp=1024$ is held constant. The saturation scale $Q_s$ for each parameter setup is estimated using the results of \cite{Lappi:2007ku} with $N_\eta=50$ color sheets in the MV model correlator from \cref{eq:MVModelSheets} and $g^2\mu L_\perp = 100$. Since we are here interested in the total $\hat{q}$ rather than directional anisotropies, we sum the momentum broadening in the $y$ and $z$ directions.

\begin{figure*}[!hbt]
\centering

    \subfigure[\,Physical units]{\includegraphics[width=0.95\columnwidth]{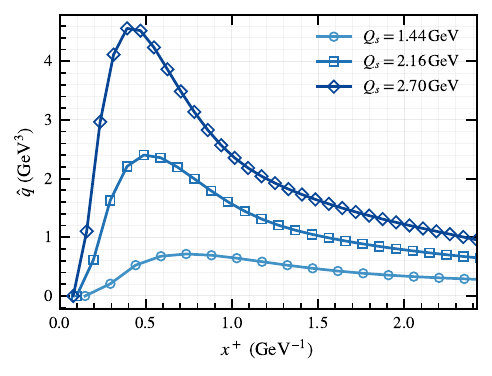} \label{fig:qhat_m=0_units}
    }
    \quad
    \subfigure[\,Dimensionless units 
    ]{\includegraphics[width=0.95\columnwidth]{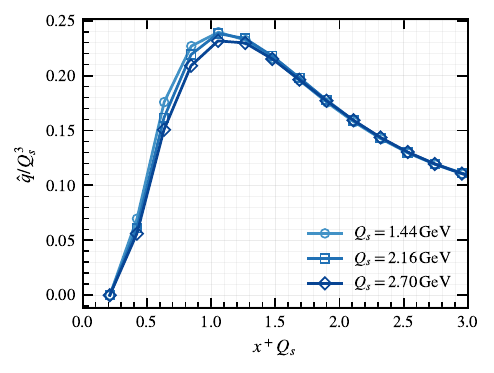} \label{fig:qhat_m=0_dimensionless}
    }
    
\caption{
Jet quenching parameter $\hat{q}$ as a function of light-cone time $x^+$ at $m_g = 0$. Different shades of blue and different markers correspond to different values of $Q_s$. Panel (a): results in physical units. Panel (b): results rescaled by $Q_s^3$ (vertical axis) and $Q_s$ (horizontal axis), showing perfect scaling with the saturation scale.
}
\label{fig:qhat_m=0}
\end{figure*}

{\boldmath\textbf{i. Case $m_g = 0$.}}
When no IR regulator is applied, the lattice resolution $\lambda_{\rm IR} = 2\pi/L_\perp$ acts as an effective IR cutoff. With $g^2\mu L_\perp = 100$, the saturation scale is related to the charge density by $Q_s/(g^2\mu) = 1.08$, giving $Q_s = 1.44$, $2.16$, and $2.70,\GeV$ for the three values of $g^2\mu$. The results are shown in \cref{fig:qhat_m=0}, both in physical units (\cref{fig:qhat_m=0_units}) and in dimensionless units (\cref{fig:qhat_m=0_dimensionless}), where $\hat{q}$ is rescaled by $Q_s^3$ and $x^+$ by $Q_s$.

All three values of $Q_s$ display the same qualitative behavior: $\hat{q}$ rises sharply at early times, reaches a peak, and then decreases as the background fields expand and dilute. This peaked behavior is consistent with previous Glasma studies \cite{Ipp:2020nfu,Carrington:2021dvw,Avramescu:2023vld}. When presented in dimensionless units (\cref{fig:qhat_m=0_dimensionless}), the curves collapse onto a single one, demonstrating a perfect scaling with $Q_s$. This is expected since, at $m_g = 0$, the saturation scale is the only dimensionful scale of the problem.

The position of the peak is located at $x^+ \sim 1/Q_s$ (equivalently $\tau \sim 1/(2Q_s)$) approximately halfway through the Glasma light-front lifetime $\Delta x^+_{\rm Glasma} \sim 2/Q_s$. This can be understood heuristically in terms of the Glasma flux tube structure \cite{Barata:2024xwy}. If the Glasma fields are organized into correlation domains of size $\sim 1/Q_s$, a jet produced at a random position will travel for an average light-front time $x^+ \sim 1/Q_s$ within a single domain before entering the next one. During this interval, the jet momentum grows linearly with time, and so does $\hat{q}$. Beyond $x^+ \sim 1/Q_s$, the jet crosses into uncorrelated domains which, after event averaging, leads to a saturation of the momentum growth. Combined with the dilution of the background fields, this produces the observed decrease of $\hat{q}$ at later times. Notably, the peak occurs while the system is still in the Glasma phase, prior to the onset of thermalization. Since $\hat{q}$ decreases with time in effective kinetic theory~\cite{Boguslavski:2024ezg}, a smooth Glasma-to-EKT matching requires $\hat{q}$ to already be decreasing at the transition point, which is indeed what we observe.

\begin{figure*}[!hbt]
\centering

    \subfigure[\,Physical units]{\includegraphics[width=0.95\columnwidth]{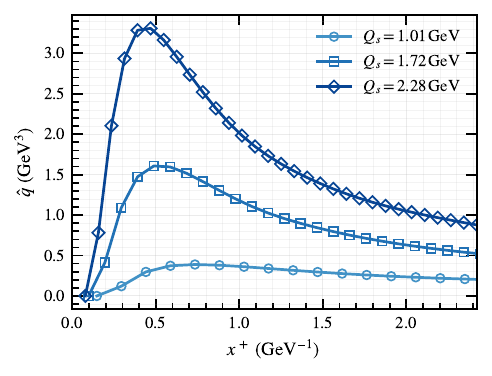} \label{fig:qhat_m=0.2_units}
    }
    \quad
    \subfigure[\,Dimensionless units]{\includegraphics[width=0.95\columnwidth]{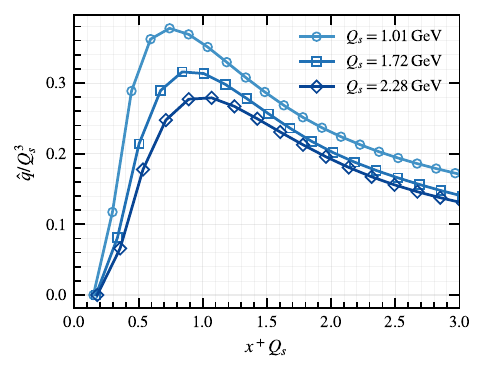} \label{fig:qhat_m=0.2_dimensionless}
    }
    
\caption{
Jet quenching parameter $\hat{q}$ as a function of light-cone time $x^+$ at $m_g = 0.2 \GeV$. Different shades of blue and different markers correspond to different values of $Q_s$. Panel (a): results in physical units. Panel (b): results rescaled by $Q_s^3$ (vertical axis) and $Q_s$ (horizontal axis), showing approximate but no longer exact scaling with $Q_s$ due to the presence of the second scale $m_g$.
}
\label{fig:qhat_m=0.2}
\end{figure*}

{\boldmath\textbf{ii. Case $m_g = 0.2\GeV$.}}
We then introduce a nonzero IR regulator $m_g = 0.2\,\GeV$, chosen such that $m_g > 2\pi/L_\perp$ and the IR physics is controlled by the gluon mass in the Poisson equation rather than by the lattice resolution. The presence of $m_g$ introduces a second dimensionful scale alongside $Q_s$, so the scaling with $Q_s$ alone is no longer expected to be exact. The ratios $m_g/(g^2\mu) = 0.15$, $0.10$, and $0.08$, obtained from \cite{Lappi:2007ku}, yield $Q_s = 1.01$, $1.72$, and $2.28,\GeV$, respectively. The results are shown in \cref{fig:qhat_m=0.2}, again in physical units in \cref{fig:qhat_m=0.2_units} and in dimensionless units in \cref{fig:qhat_m=0.2_dimensionless}. Comparing \cref{fig:qhat_m=0_dimensionless} and \cref{fig:qhat_m=0.2_dimensionless}, the presence of the IR regulator shifts the peak to earlier times and increases its magnitude in dimensionless units. However, the dependence on $m_g$ remains weaker than the leading scaling behavior given by the dependence on $Q_s$. Both effects are in good agreement with the predictions of Ref. \cite{Ipp:2020nfu}.

\subsubsection{Phenomenological estimates}

The obtained values of $\hat{q}$ can be used to parametrically estimate the energy loss during the Glasma phase and to compare to the energy loss in the QGP phase. We here recall that the BDMPS-Z picture of independent soft scatterings is not well suited for the Glasma, so the results in this section must be understood as an educated estimate and not a precise numerical result.

The length traveled by the quark inside the Glasma phase can be estimated as $L_{Glasma} \sim 1/Q_s$. Integrating the obtained values of $\hat{q}$ up to this life-time the momentum broadening accumulated during the Glasma phase is found to be $\langle (\delta p_\perp^{kin})^2\rangle_{\mathrm{Glasma}} \approx 0.14 Q_s^2$ for $m_g = 0 \GeV$. For $m_g = 0.2 \GeV$ the value is slightly larger, but the difference is nevertheless negligible at parametric level. The average value of $\hat{q}$ during the Glasma phase is therefore $\hat{q}_{\mathrm{Glasma}} \sim 0.14 Q_s^3$. 
The value of the jet quenching parameter in the QGP extracted from data -- see \cite{Apolinario:2022vzg} for a recent review containing many phenomenological analyses -- is rather unconstrained, in the range $2\lesssim \hat q/T^3\lesssim 5$. As we just need a reasonable estimate for our comparisons, we will take $\hat{q}_{\mathrm{QGP}} \sim 3 T^3$. For Pb-Pb collisions, using typical values $Q_{s,Pb} \approx 2 \GeV$ and $T_{Pb} \approx 0.2 \GeV$ we find
\begin{align} \label{eq:qhatRatio}
    \frac{\hat{q}_{\mathrm{Glasma}}}{\hat{q}_{\mathrm{QGP}}} \approx 50\, ,
\end{align}
so the jet quenching parameter in the Glasma phase is between one and two orders of magnitude larger than that on the QGP. As the dominant scale 
in the problem is the saturation scale $Q_s$ of the colliding nuclei, we can assume parametrically that $T \propto Q_s$ when varying the atomic mass number $A$ or the collision energy. The ratio in \cref{eq:qhatRatio} will then hold over different nuclear species.

Let us now see how this large value of $\hat{q}$ translates into medium-induced momentum broadening and energy loss. As we previously exposed, the Glasma is a valid description for a path-length $L_{\mathrm{Glasma}} \sim 1/Q_s$. We also know that $Q_s$ scales with the nuclear mass number as $Q_s^2 \propto A^{1/3}$ \cite{McLerran:1993ni}, so the saturation scale for any nuclear species can be estimated from the one of lead collisions through $Q_{s} \sim (A/A_{\mathrm{Pb}})^{1/6} Q_{s,\mathrm{Pb}}$. In addition, we consider that the QGP size traversed by the jet is, on average, the nuclear radius, $L_{\mathrm{QGP}} \sim R_\mathrm{A} \sim 1.2 \fm \cdot A^{1/3}$. The ratio between the momentum broadening accumulated in the Glasma and the QGP is then
\begin{align}\label{eq:MomBroadRatio}
    \frac{\langle (\delta p_\perp^{kin})^2\rangle_{\mathrm{Glasma}}}{\langle (\delta p_\perp^{kin})^2\rangle_{\mathrm{QGP}}} = \frac{\hat{q}_{\mathrm{Glasma}}}{\hat{q}_{\mathrm{QGP}}} \frac{L_{\mathrm{Glasma}}}{L_{\mathrm{QGP}}} \sim 50\, \frac{(A_{\mathrm{Pb}}/A)^{1/6}}{12A^{1/3}}\, ,
\end{align}
which for Pb-Pb and O-O collisions yields 
\begin{align}\label{eq:MomBroadRatioResult}
    \left. \frac{\langle (\delta p_\perp^{kin})^2\rangle_{\mathrm{Glasma}}}{\langle (\delta p_\perp^{kin})^2\rangle_{\mathrm{QGP}}} \right|_{\mathrm{Pb}} \approx 0.7\, , \quad \left. \frac{\langle (\delta p_\perp^{kin})^2\rangle_{\mathrm{Glasma}}}{\langle (\delta p_\perp^{kin})^2\rangle_{\mathrm{QGP}}} \right|_{\mathrm{O}} \approx 2.5\, .
\end{align}
The momentum broadening induced on jets by the Glasma nuclear matter is comparable to the one induced by the QGP even for large systems, consistent with similar observations in Ref. \cite{Carrington:2021dvw}. In light nuclei collisions the thermalization time increases while the QGP lifetime decreases, therefore yielding a much larger relative importance of the Glasma stage over the thermalized stage. In this kind of small systems the contribution of the Glasma phase to momentum broadening can even be larger than the QGP contribution.

The increase in the  relevance of the Glasma phase for small collision systems is even more evident when one looks at medium induced energy loss.
The ratio between the energy loss induced by the Glasma and that by the QGP follows from \cref{eq:EnergyLossEstimate} as
\begin{align}\label{eq:EnergyLossratio}
    \frac{\Delta E_{\mathrm{Glasma}}}{\Delta E_{\mathrm{QGP}}} = \frac{\hat{q}_{\mathrm{Glasma}}}{\hat{q}_{\mathrm{QGP}}} \left(\frac{L_{\mathrm{Glasma}}}{L_{\mathrm{QGP}}}\right)^2 \sim 50 \frac{(A_{\mathrm{Pb}}/A_{\mathrm{X}})^{1/3}}{144 A_{\mathrm{X}}^{2/3}}\, ,
\end{align}
which scales with the squared medium length, so that the relative Glasma contribution will be largely enhanced in small collision systems. For Pb-Pb and O-O collisions this estimate yields
\begin{align}\label{eq:EnergyLossRatioResult}
    \left. \frac{\Delta E_{\mathrm{Glasma}}}{\Delta E_{\mathrm{QGP}}} \right|_{\mathrm{Pb}} \approx 0.01\, , \qquad \left. \frac{\Delta E_{\mathrm{Glasma}}}{\Delta E_{\mathrm{QGP}}} \right|_{\mathrm{O}} \approx 0.13\, .
\end{align}
The energy loss in the Glasma phase is therefore largely suppressed compared to the QGP one in heavy-ion collisions due to the longer lifetime of the QGP phase. This can make  it difficult to experimentally access the Glasma stage. However, in light-ion collisions the Glasma-induced energy loss becomes sizable, only one order of magnitude smaller than the one induced by the QGP. Consequently, we believe that the ongoing light-ion collision program at the LHC \cite{Brewer:2021kiv, CMS:2025bta, Citron:2018lsq} can offer a unique opportunity to probe the effects of the Glasma on jet observables.

Finally, we stress that the energy loss estimate above relies on the assumption of independent soft scatterings, which is not strictly valid in the Glasma due to the presence of large field correlations in the form of flux tubes and correlation domains, as discussed in \cref{sec:qhat_calc}. A rigorous computation of Glasma-induced energy loss requires including at least the $\ket{qg}$ sector of the quark Fock space and tracking the gluon emission probability over time. We leave this full calculation for future work. 

\subsection{The color rotation}

Let us finally study the color rotation experienced by the quark quantum state through  its interaction with the Glasma background field. 
Although the color is not a physical observable, analyzing its evolution provides valuable information about the mechanisms of in-medium jet evolution and the intensity of the background fields. We consider jets initially prepared in a single color state $c=1$, and evolve the wavefunction in time. We then compute the probability of finding the quark in each of the color states $c = 1, 2, 3$ as a function of time, according to \cref{eq:ColorEvolution}. This probability depends directly on the magnitude of the longitudinal component of the background field $\mathcal{A}_+$ and is therefore a gauge-dependent quantity. It is also sensitive to the MV model parameters, namely the saturation scale $Q_s$ and the IR regulator $m_g$.

\subsubsection{Gauge dependence}

We begin by studying how the color rotation depends on the gauge choice. 
The Glasma fields are derived in the temporal gauge $\mathcal{A}_\tau = 0$. In all previous sections, we have used the residual gauge freedom to additionally impose the transverse Coulomb gauge condition $\partial_{\underline{i}} \mathcal{A}_{\underline{i}} = 0$, with $\underline{i} \in \{x,y\}$, as this choice reduces the magnitude of the background fields and thereby suppresses numerical errors \cite{Avramescu:2026fgv}. Here, we present results for the color rotation both with and without this additional condition, in order to assess the effect of the gauge transformation on the jet wavefunction. We chose the charge density and the effective gluon mass to be $g^2\mu = 2 \GeV$ and $m_g = 0$. We refer to the case where the residual gauge freedom in temporal gauge is supplemented by fixing the Coulomb gauge condition at the initial time as the \textit{Coulomb gauge}, and to the case without the Coulomb gauge fixing as the \textit{temporal gauge}.

\begin{figure}[!hbt]
\centering
    \includegraphics[width=0.95\columnwidth]{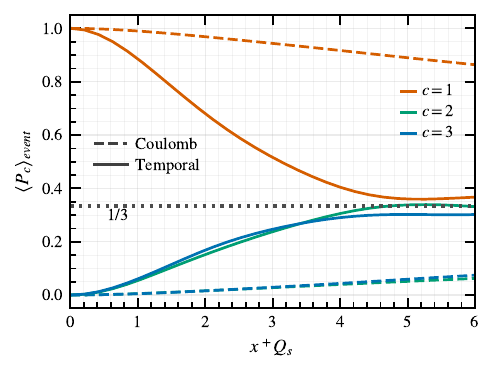}    
\caption{Gauge dependence of the color rotation. The probability of finding the quark state in a $c=1$ (red line), $c=2$ (green line) and $c=3$ (blue line) state is shown as a function of dimensionless time $x^+Q_s$. Results are shown both before (solid line) and after (dashed line) imposing the Coulomb gauge condition on the background fields. 
}
\label{fig:ColorRotGauge}
\end{figure}

Results are shown in \cref{fig:ColorRotGauge} where we find the color rotation to be much faster in the temporal gauge than in the Coulomb gauge. This is an expected result as the Coulomb gauge choice is such that it minimizes the surface generated by the $\mathcal{A}_x$ and $\mathcal{A}_y$ components of the background field. The color rotation depends directly on the longitudinal $\mathcal{A}_+$ component so, by reducing its magnitude, the color homogenization will become slower. As explained in \cite{Avramescu:2026fgv}, a rapid color rotation induced by a large $\mathcal{A}_x$ is a potential source of numerical error accumulation. Therefore, the Coulomb gauge improves the robustness of our numerical scheme, since the color rotation is almost negligible in the Glasma phase ($x^+ \lesssim 2/Q_s)$. We believe the small deviation from the symmetry value $1/3$ at late times to be a statistical fluctuation. Even with rather large statistics in the configuration average (with $100$ Glasma events), the large pure gauge component present in the temporal gauge fields has large  fluctuations, which we have confirmed by looking at the statistical error (not shown on the plot). 

\subsubsection{Saturation scale dependence}

\begin{figure}[!tb]
\centering
    \includegraphics[width=0.95\columnwidth]{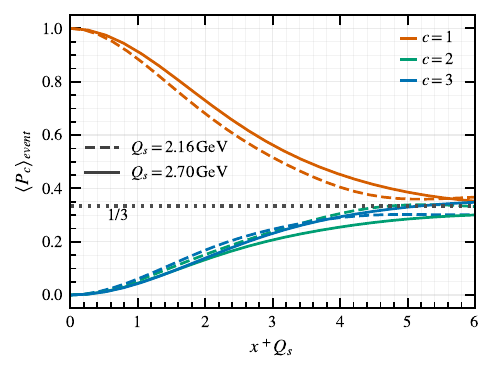}    
\caption{Dependence of the color rotation on the saturation scale. The probability of finding the quark state in a $c=1$ (red line), $c=2$ (green line) and $c=3$ (blue line) state is shown as a function of dimensionless time $x^+Q_s$. Results are obtained in the temporal gauge, without the imposition of the Coulomb gauge condition, for $Q_s = 2.16 \GeV$ (dashed line) and $Q_s = 2.70 \GeV$ (solid line).
}
\label{fig:ColorRotg2mu}
\end{figure}

We next examine how the color rotation depends on the Glasma model parameters, in particular on the saturation scale $Q_s$. We work in the temporal gauge, where the color rotation is more pronounced and therefore more sensitive to the background field parameters. With $m_g = 0$, the saturation scale is the only dimensionful scale of the background fields. We take $g^2\mu = 2\,\GeV$ and $g^2\mu = 2.5\,\GeV$, corresponding to $Q_s = 2.16\,\GeV$ and $Q_s = 2.70\,\GeV$, respectively. Results are shown in \cref{fig:ColorRotg2mu}. When plotted as a function of the dimensionless time $x^+ Q_s$, the two datasets exhibit near-perfect scaling, consistent with $Q_s$ being the only relevant scale. Residual deviations are of the same order as the difference between the $c=2$ and $c=3$ components and are therefore attributed to large statistical fluctuations associated with working without the Coulomb gauge fixing.

\section{Conclusion and outlook}\label{sec:summary}

We studied the real-time quantum evolution of a high-energy quark jet propagating through the Glasma phase of a heavy-ion collision using the light-front Hamiltonian formalism and time-dependent basis light-front quantization (tBLFQ). We derived the light-front Hamiltonian for quark propagation in an external background field and solved its real-time evolution in the eikonal limit, where the time evolution of the quark state is driven by the longitudinal potential $V_I=2g\mathcal A_+$. The transverse gauge fields $\vec{\mathcal{A}}_\perp$ also enter our measurement through the kinetic momentum operator. The quark quantum state is expanded in a discrete basis representation spanned by transverse momentum, longitudinal momentum, and color, and its Fock space is truncated to the $\ket{q}$ sector.
The real-time evolution of the quark state is then obtained by solving the time-dependent Schrödinger equation on the light front, with the background Glasma fields entering as a time-dependent external potential evaluated on a classical real-time lattice.

Our main findings are the following:
\begin{itemize}
    \item We computed transverse momentum broadening $(\delta p^{kin}_i)^2$ via two different quantum methods. 
    The first evaluates the kinetic momentum operator directly using the time-evolved quark state in the Schrödinger picture. The second works in the Heisenberg picture, computing the time derivative of $\vec{p}^{\,kin}_\perp$ to obtain the quantum Lorentz force operator, which is then integrated and squared to yield $(\delta p^{kin}_i)^2$. This second method is the quantum analogue of integrating Wong's equations in the classical calculation of Ref.~\cite{Ipp:2020mjc,Avramescu:2023qvv,Avramescu:2026fgv}, with forces and fields understood as quantum operators rather than c-numbers. Both methods agree with each other and  with the classical result of Ref.~\cite{Ipp:2020mjc,Avramescu:2023qvv}, validating our formalism within the current Fock space truncation where gluon emissions are absent.

    \item We studied the dependence of momentum broadening on the transverse width of the jet wave packet. While $(\delta p^{kin}_y)^2$ is insensitive to the spread $\sigma_y$, it increases with $\sigma_z$ as the delocalized jet probes Glasma fields at transverse positions away from $z=0$, where they differ from their values at the collision point. 
    Conversely, $(\delta p^{kin}_z)^2$ is insensitive to both $\sigma_y$ and $\sigma_z$. This is not a purely quantum effect, but rather a consequence of wave-packet delocalization, and we expect the same behavior for a classical ensemble of probe particles distributed in the transverse plane. We note that our coordinate matching between Glasma and light-cone frames relies on approximations valid only for a well-localized jet around $z=0$, so these delocalization results should be interpreted as qualitative predictions.

    \item We obtained the jet quenching parameter $\hat{q}$ as a function of light-cone time for different values of $Q_s$. For $m_g=0$ it peaks at $x^+ \approx 1/Q_s$ (or $\tau \approx 1/(2Q_s)$) and we find a perfect scaling with the saturation scale $\hat{q}/Q_s^3 = \mathrm{const}$. When a finite IR regulator $m_g = 0.2 \GeV$ is introduced, the exact scaling with $Q_s$ is somewhat violated:  the peak shifts to earlier times and its magnitude increases in dimensionless units, in good agreement with the predictions of Ref.~\cite{Ipp:2020nfu}.

    \item We investigated the color rotation of the quark state induced by propagation through the Glasma, which serves as a direct probe of the  interaction potential $V_I = 2g\mathcal{A}_+$. The isotropization times scales as $1/Q_s$ when $m_g = 0$. Imposing the Coulomb gauge condition on the  background fields significantly slows the color rotation, suppressing the accumulation of numerical errors in the time evolution \cite{Avramescu:2026fgv}.
\end{itemize}

The present work establishes the tBLFQ framework as a viable tool for studying real-time quantum jet evolution in the Glasma. A phenomenological comparison with QGP estimates suggests that Glasma-induced effects can be most prominent in light-ion collisions such as O-O \cite{Brewer:2021kiv, CMS:2025bta}, where the Glasma phase constitutes a larger fraction of the total evolution time. Our work provides the starting point for going beyond the $|q\rangle$ Fock sector truncation and include the $|qg\rangle$ sector. This will enable the explicit computation of medium-induced gluon emission and radiative energy loss, providing a more complete quantum picture of jet quenching in the Glasma. 
Additionally, this formalism can be extended to study the quantum evolution of a gluon probe. At the considered eikonal order the extension is straightforward: one replaces the fundamental color generators with the adjoint ones and the momentum broadening is expected to scale with the Casimir factor. Casimir scaling of classical quark vs. gluon momentum broadening in the Glasma was studied in \cite{Ipp:2020mjc, Avramescu:2023qvv} where it was found to hold for eikonal and static quarks, and starts to break for classical particles with finite $m$ or $p^+$. Attempts to use the light-front Hamiltonian formalism to evolve a gluon probe have been made in the QGP and cold-nuclear matter \cite{Qian:2024gph, Castro:2025ocx}. The gluon-probe extension in the Glasma is part of our plans for future work.

\section*{Acknowledgments}
We are grateful to Jo\~{a}o Barata, Yang Li, Xo\'{a}n Mayo L\'{o}pez, David M\"{u}ller, and Wenyang Qian for their helpful and valuable discussions.

DA and TL are supported by the Research Council of Finland, the Centre of Excellence in Quark Matter (projects 346324 and 364191). DA also acknowledges the support of the Vilho, Yrj\"{o} and Kalle V\"{a}is\"{a}l\"{a} Foundation. CL, ML, and CS are supported by European Research Council under project ERC-2018-ADG-835105 YoctoLHC; by Maria de Maeztu excellence unit grant CEX2023-001318-M and project PID2023-152762NB-I00 funded by MICIU/AEI/10.13039/501100011033; by ERDF/EU; and by Xunta de Galicia (CIGUS Network of Research Centres). CL also acknowledges support from the Ministerio de Ciencia e Innovación through the predoctoral fellowship PRE2022-102748 funded by MCIN/AEI/10.13039/501100011033.
TL acknowledges support from grant NSF PHY-2309135 to the Kavli Institute for Theoretical Physics (KITP). ML and CL are also supported by Xunta de Galicia under Project ED431F 2023/10.

\appendix

\section{Coordinate conventions} \label{app:Coordinate_Conventions}

In this work, we employ two distinct coordinate systems. 
The first consists of Milne coordinates, $x^\mu_{(M)}$, which are natural for the time evolution of the boost-invariant Glasma.
The second uses the light-cone coordinates defined along the jet direction, $x^\mu$, which are well suited for the time evolution of the jet. In addition, to obtain the Glasma initial condition in \cref{sec:GlasmaFields} we use the light-cone coordinates referred to the beam direction $x^{\bullet *}$.

We here summarize the conventions adopted for those coordinate systems and describe their relations to the standard Minkowski coordinates.

\subsection{Milne coordinates} \label{app:MilneCoordinates} 

The Milne coordinates are $x^\mu_{(M)} = (\tau, x, y, \eta)$, where $\tau$ is the proper time and $\eta$ the space-time rapidity. They are convenient for describing the evolution of the boost-invariant Glasma. They are related to the Minkowski coordinates by
\begin{align} \label{eq:MilnetoMinkowski}
    \tau = \sqrt{t^2-z^2}\, , \quad \eta = \tanh^{-1}(z/t)\, ,
\end{align}
or the inverse relation
\begin{equation} \label{eq:MinkowskitoMilne}
    t = \tau \cosh\eta \, , \quad z = \tau \sinh\eta\, .
\end{equation}

The metric tensor can be obtained through
\begin{align} \label{eq:MetricRelation}
    g^{(M)}_{\mu\nu}
    = \frac{\partial x^\alpha_{(M)}}{\partial x^\mu} \frac{\partial x^\beta_{(M)}}{\partial x^\nu} g_{\alpha\beta}\, ,
\end{align}
in which $g_{\mu\nu}=\mathrm{diag}(1, -1, -1, -1)$ is the metric of the Minkowski coordinates. Using the coordinate transformation in \cref{eq:MilnetoMinkowski} the Milne coordinate metric tensor reads
\begin{align}
    g_{\mu\nu}^{(M)} = (1, -1, -1, -\tau^2) \, , \quad
    g^{\mu\nu}_{(M)} = (1, -1, -1, -1/\tau^2)\, .
\end{align}
In this set of curvilinear coordinates, it is important to differentiate between upper and lower indices, related through the metric tensor
\begin{align} \label{eq:MilneVectorTransformation}
    V_\tau = V^\tau\, , \quad V_x = -V^x \, , \quad V_y = -V^y \, , \quad V_\eta = -\tau^2 V^\eta \, .
\end{align}

\subsection{Jet light-cone coordinates} \label{app:JetLCCoordinates}

We define the light-cone coordinates of the jet as $x^\mu = (x^+, x^-, y, z)$, where
\begin{align} \label{eq:LCtoMinkowski}
    x^\pm = t \pm x\;.
\end{align}
In our setup, the jet propagates ultra-relativistically along the positive $x$-direction. Accordingly, its light-front time is identified with $x^+$, while $x^-$ serves as the longitudinal spatial direction. 

In analogy to using \cref{eq:MetricRelation}, we find the metric of the light-cone coordinates as
\begin{align} \label{eq:LC}
    g_{\mu\nu}^{(LC)} = \begin{pmatrix}
        0 & \tfrac{1}{2} & 0 & 0 \\
        \tfrac{1}{2} & 0 & 0 & 0 \\
        0 & 0 & -1 & 0 \\
        0 & 0 & 0 & -1
    \end{pmatrix}\, , \quad g_{(LC)}^{\mu\nu} = \begin{pmatrix}
        0 & 2 & 0 & 0 \\
        2 & 0 & 0 & 0 \\
        0 & 0 & -1 & 0 \\
        0 & 0 & 0 & -1
    \end{pmatrix}\, .
\end{align}
In these coordinates, vectors then transform as
\begin{align} \label{eq:LCVectorTransformation}
    V_+ = \frac{1}{2} V^- \, , \quad V_- = \frac{1}{2}V^+ \, , \quad V_y = -V^y \, , \quad V_z = -V^z \, ,
\end{align}
and the scalar product of two vectors is given by
\begin{equation}
    \begin{split}
        V^\mu W_\mu & = g_{\mu\nu}^{(LC)} V^\mu W^\nu \\
    & = \frac{1}{2} (V^+ W^- + V^- W^+) -V^y W^y - V^z W^z\, .
    \end{split}
\end{equation}

\subsection{Beam light-cone coordinates} \label{app:NucleiLCCoordinates}
We define the light-cone coordinates of the nuclei, propagating along $z$ direction, as $x^{\bullet *} = (x^\bullet, x, y, x^*)$, where
\begin{align}
    x^\bullet = t + z \, \qquad x^* = t - z\,.
\end{align}
The metric tensor and vector transformations are identical to those in the jet light-cone coordinate system under the $x \leftrightarrow z$ exchange.

\section{Quantum many-body representation of the light-front Hamiltonian}
\label{app:HamiltonianSimplification}

In this section, we provide a derivation of the quantum many-body representation of the light-front Hamiltonian given in \cref{eq:1stQuantHamiltonian} from the light-front quantized Hamiltonian \cref{eq:LFHamiltonianSubEik}. The derivation closely follows the framework introduced in Ref.~\cite{Wen:2025dwy}.
    
{\boldmath\textbf{i. Vacuum kinetic energy.}} The first term in \cref{eq:LFHamiltonianSubEik} is the kinetic energy of a fermion propagating in vacuum, and it can be written in a second-quantized form by substituting the free field expansion given in \cref{eq:free_mode_expansion}. Here, we keep the quark part, and drop off the anti-quark component 
\begin{align} \label{eq:VacuumKinetic}
    \begin{split}
        T_q & = \int \diff x^- \diff^2\vec{x}_\perp 
        \frac{1}{2} \bar{\Psi}\gamma^+ 
        \frac{m^2-\nabla_\perp^2}{i\partial^+} \Psi \\
        & =
        \sum_{\lambda,c}
        \int\diff p^+ \diff^2 x_\perp \int \frac{\diff^2 p_\perp \diff^2 q_\perp}{(2\pi)^2} e^{i \vec{x}_\perp \cdot (\vec{p}_\perp - \vec{q}_\perp)} \\
        & \hspace{40 pt} \times b_{\lambda, c}^\dagger(p^+, \vec{q}_\perp)
        \frac{m^2+p_\perp^2}{p^+} b_{\lambda, c} (p^+, \vec{p}_\perp)\, ,
    \end{split}
\end{align}
where we have used the spinor matrix element
\begin{align} \label{eq:PlusQuarkSPinor}
    \bar{u}_{\lambda^\prime}(p) \gamma^+ u_{\lambda}(p') = 2p^+ \delta_{\lambda \lambda^\prime} \, .
\end{align}
One can now integrate out $\vec{x}_\perp$, fixing $T_q$ to be diagonal in momentum space. Alternatively, we can write the kinetic energy in the transverse position space. For this, we introduce the transverse-position-space representation by defining
\begin{align}
\begin{split}
   & B_{\lambda,c}(p^+, \vec x_\perp)\equiv \frac{1}{2\pi} \int\diff^2 p_\perp e^{i\vec p_\perp\cdot \vec x_\perp} b_{\lambda,c}(p)\;,\\
   &    b_{\lambda,c}(p) = \frac{1}{2\pi} \int\diff^2 x_\perp e^{-i\vec p_\perp\cdot \vec x_\perp} B_{\lambda,c}(p^+, \vec x_\perp)\;,
\end{split}
\end{align}
and likewise for $B^\dagger$. It follows that
\begin{align}
    \begin{split}
       T_q =&
        \sum_{\lambda,c}
        \int \diff^2 x_\perp
        \diff p^+ \\
        & \times B^\dagger_{\lambda,c}(p^+, \vec x_\perp)
        \frac{m^2+(i\nabla_\perp )^2}{p^+} B_{\lambda,c}(p^+, \vec x_\perp)\\
    \end{split}
\end{align}
Note that the canonical momentum operator in coordinate space is $\vec p_\perp= i\,\vec \nabla_\perp $. 

{\boldmath\textbf{ii. Interaction potential.}}
The term in the Hamiltonian containing $\mathcal{A}_+$ represents the time-dependent interaction potential in our calculation. Using the same spinor relation as for the vacuum kinetic energy term it follows that
\begin{align} \label{eq:LogintudinalInteraction}
    \begin{split}
        V_{\mathcal{A}_+} = &\int \diff x^- \diff^2\vec{x}_\perp 
        g \bar{\Psi} \gamma^+ \mathcal{A}_+ 
        \Psi \\
         = &\sum_{\lambda} \sum_{c, c^\prime} \int \diff p^+ \diff^2 \vec{x}_\perp B^\dagger_{\lambda c^\prime} (p^+, \vec{x}_\perp) \\
        & \hspace{30 pt} \times 2g\mathcal{A}_{+, c'c}(x^+, \vec{x}_\perp) B_{\lambda, c}(p^+, \vec{x}_\perp) \, .
    \end{split}
\end{align}

{\boldmath\textbf{iii. Interaction with $\vec{\mathcal{A}}_i$.}}
Let us now consider the interaction with the field components transverse to the jet direction
\begin{align} \label{eq:TransverseInteraction}
    \begin{split}
        V_{\vec{\mathcal{A}}_\perp} & = \int \diff x^- \diff^2\vec{x}_\perp g \bar{\Psi} \gamma^i \mathcal{A}_i \Psi \\
        & = \sum_{\lambda, \lambda^\prime} \sum_{c, c^\prime} \int \diff p^+ \diff^2 \vec{x}_\perp B^\dagger_{\lambda^\prime, c^\prime} (x^+, \vec{x}_\perp) \\
        & \hspace{30 pt} \times g \left[ - \frac{(i\vec{\nabla}_\perp) \cdot \vec{\mathcal{A}}_{\perp, cc^\prime} + \vec{\mathcal{A}}_{\perp, cc^\prime} \cdot (i\vec{\nabla}_\perp)}{p^+} \delta_{\lambda \lambda^\prime} \right. \\
        & \hspace{30 pt} \left. + \frac{(\vec{\nabla}_\perp \times \vec{\mathcal{A}}_{\perp, c^\prime c}) \cdot \vec{\sigma}_{\lambda \lambda^\prime}}{p^+}  \right] B_{\lambda, c}(x^+, \vec{x}_\perp) \, ,
    \end{split}
\end{align}
where we have used the spinor relation 
\begin{align}
    \bar{u}_{s^\prime}(q)\gamma^iu_s(p) \mathcal{A}_i &= -(\vec{p}_\perp + \vec{q}_\perp) \cdot \vec{\mathcal{A}}_\perp \delta_{\lambda, \lambda^\prime} \notag \\
    &- i [(\vec{q}_\perp - \vec{p}_\perp) \times \vec{\mathcal{A}}_\perp] \cdot \vec{\sigma}_{\lambda' \lambda}\;,
\end{align}
and $\vec \sigma=\{\sigma^1, \sigma^2, \sigma^3\}$ are the Pauli matrices corresponding to the $y$, $z$, and $x$ components in the defined coordinate system, respectively. 

{\boldmath\textbf{iv. Instantaneous interaction.}}
By last, let us consider the last term in \cref{eq:LFHamiltonianSubEik}
\begin{align} \label{eq:InstantaneousInteraction}
    \begin{split}
        W_{q} = &\int \diff x^- \diff^2\vec{x}_\perp 
        \frac{g^2}{2} \bar{\Psi} \gamma^i \mathcal{A}_i \frac{\gamma^+}{i\partial^+} \gamma^j \mathcal{A}_j \Psi \\
        = & g^2\sum_{\lambda, \lambda^\prime} \sum_{c, c^\prime} \int \diff p^+ \diff^2 \vec{x}_\perp B^\dagger_{\lambda',c'}(p^+, \vec{x}_\perp) \\
        & \hspace{10 pt} \times  \frac{(\mathcal{A}_\perp^2 \delta_{\lambda \lambda^\prime} + i \sigma^3_{\lambda, \lambda^\prime}[\mathcal{A}_y, \mathcal{A}_z])_{c' c}}{p^+} B_{\lambda,c}(p^+, \vec{x}_\perp) \, ,
    \end{split}
\end{align}
where we have used the spinor relation
\begin{align}
    \bar{u}_{\lambda^\prime}(q) \gamma^i &\gamma^+ \gamma^j u_\lambda(p) \mathcal{A}_i \mathcal{A}_j \notag \\
    &= 2\sqrt{p^+ q^+} (\delta_{\lambda, \lambda^\prime} \mathcal{A}_\perp^2 + i \sigma^3_{\lambda', \lambda} [\mathcal{A}_y, \mathcal{A}_z])\, .
\end{align}

The first term in $V_{\vec{\mathcal{A}}_\perp}$ and that in the instantaneous interaction $W_{q}$ combine together with the kinetic term to give the kinetic energy of the particle propagating inside the medium $(\vec{p}_\perp - g\vec{\mathcal{A}}_\perp)^2/p^+$. 
The second term in $V_{\vec{\mathcal{A}}_\perp}$ and that in $W_{q}$ combine together to yield $- \mathcal{B}_x = \mathcal{F}_{yz} = \partial_y \mathcal{A}_z - \partial_z \mathcal{A}_y + ig[\mathcal{A}_y, \mathcal{A}_z]$.
The full light-front Hamiltonian can then be expressed as
\begin{widetext}
\begin{align}\label{eq:LFH_full_many_body}
    \begin{split}
    P^- = T_q + V_{\mathcal{A}^+} + V_{\vec{\mathcal{A}}_\perp} + W_q & = \sum_{\lambda, \lambda^\prime} \sum_{c, c^\prime} \int \diff p^+ \diff \vec{x}_\perp B^\dagger_{\lambda^\prime, c^\prime}(p^+, \vec{x}_\perp) \\
    & \times \left[ \frac{m^2 \delta_{c, c^\prime} + (\vec{p}_\perp \delta_{c, c^\prime} - g \vec{\mathcal{A}}_{\perp, c^\prime c})^2}{p^+} \delta_{\lambda, \lambda^\prime} + 2g \mathcal{A}_{+, c^\prime c} \delta_{\lambda, \lambda^\prime} - g \frac{\mathcal{B}_{x, c^\prime c} \sigma^3_{\lambda', \lambda}}{p^+} \right] B_{\lambda, c}(p^+, \vec{x}_\perp) \;.
    \end{split}
\end{align}
which can be written in the more compact form of 
\cref{eq:1stQuantHamiltonian} in the main text.
\end{widetext}

\section{The Lorentz force quantum operator} \label{app:Lorentz}
We present here the derivation of the Lorentz force operator in \cref{eq:LorentzForceOperator} from quantum first principles. The Heisenberg equation of motion for the kinetic momentum operator $\vec{p}^{\,kin}_\perp = \vec{p}_\perp - g \vec{\mathcal{A}}_\perp$ is
\begin{align}
    \frac{\diff p_{i,H}^{kin}}{\diff x^+} = i \left[ \frac{1}{2}P^-_H, p^{kin}_{i,H} \right] + \frac{\partial p_{i,H}^{kin}}{\partial x^+}\, .
\end{align}
Since the canonical momentum operator $p_i$ has no explicit light-front time dependence, the partial derivative term receives contributions only from the background field
\begin{align}
    \frac{\partial p_{i,H}^{kin}}{\partial x^+} = -g\,U^\dagger(x^+; 0)\, \partial_+ \mathcal{A}_i \,U(x^+; 0)\, .
\end{align}
The commutator of the canonical momentum with the eikonal Hamiltonian in \cref{eq:EikonalHamiltonian} gives 
\begin{align}
    i \left[ \frac{1}{2}P^-_H, p_{i,H} \right] = g\, U^\dagger(x^+; 0) \,\partial_i \mathcal{A}_+\, U(x^+; 0)\, ,
\end{align}
where we used the identity $[p_i, f] = i\partial_i f$ for any function $f$.  Finally, the commutator involving the background field yields 
\begin{align}
    i \left[ \frac{1}{2}P^-_H, \mathcal{A}_{i,H} \right] = ig^2 U^\dagger(x^+, 0) [\mathcal{A}_+, \mathcal{A}_i]U(x^+,0) \, .
\end{align}
Collecting all three contributions, the Heisenberg picture Lorentz force operator $f_{i,H}$ takes the form
\begin{align}
        f_{i, H}(x^+) = 2\, U^\dagger(x^+; 0) \mathcal{F}_{i+} U(x^+; 0) \, ,
\end{align}
where $\mathcal{F}_{i+}=\partial_i \mathcal{A}_+ - \partial_+ \mathcal{A}_i - ig[\mathcal{A}_+, \mathcal{A}_i]$ is the field strength tensor of the background field. Transforming back to the Schr\"{o}dinger picture recovers the result in \cref{eq:LorentzForceOperator}. Note that this is a first principle quantum calculation and $f_i$ must therefore be understood as an operator acting over a quantum state. 

\bibliography{papers/paper}

\end{document}